%% file: Main.tex
\documentclass[12pt,draftclsnofoot,onecolumn]{IEEEtran}
\IEEEoverridecommandlockouts

\usepackage{graphicx}
\usepackage{subfigure}
\usepackage{amsmath,amsthm,amsfonts,amssymb}
\usepackage{cite}
\usepackage{bm}
\usepackage{url}
\usepackage{array}
\usepackage{color}
\usepackage{multirow}
\usepackage{booktabs}
\usepackage[table,xcdraw]{xcolor}

\theoremstyle{plain}
\newtheorem{thm}{Theorem}

\newtheorem{prop}[thm]{Proposition}

\newenvironment{NewProof}{{\noindent\it Proof.}}{\hfill $\blacksquare$\par}

\allowdisplaybreaks[4]

\makeatletter
\def\old@comma{,}
\catcode`\,=13
\def,{%
  \ifmmode%
    \old@comma\discretionary{}{}{}%
  \else%
    \old@comma%
  \fi%
}
\makeatother

\begin{document}
\title{New Transceiver Designs for Interleaved Frequency Division Multiple Access}

\author{
Soung~Chang~Liew,~\IEEEmembership{Fellow,~IEEE},
Yulin~Shao,~\IEEEmembership{Student Member,~IEEE}
\thanks{S. C. Liew and Y. Shao are with the Department of Information Engineering, The Chinese University of Hong Kong, Shatin, New Territories, Hong Kong (e-mail: \{soung, sy016\}@ie.cuhk.edu.hk).}
}

\maketitle

\begin{abstract}
This paper puts forth a class of new transceiver designs for interleaved frequency division multiple access (IFDMA) systems.
These transceivers are significantly less complex than conventional IFDMA transceiver.
The simple new designs are founded on a key observation that multiplexing and demultiplexing of IFDMA data streams of different sizes are coincident with the IFFTs and FFTs of different sizes embedded within the Cooley-Tukey recursive FFT decomposition scheme.
For flexible resource allocation, this paper puts forth a new IFDMA resource allocation framework called Multi-IFDMA, in which a user can be allocated multiple IFDMA streams.
Our new transceivers are unified designs in that they can be used in conventional IFDMA as well as multi-IFDMA systems.
Two other well-known multiple-access schemes are localized FDMA (LFDMA) and orthogonal FDMA (OFDMA).
In terms of flexibility in resource allocation, Multi-IFDMA, LFDMA, and OFDMA are on an equal footing.
With our new transceiver designs, however, IFDMA has the following advantages (besides other known advantages not due to our new transceiver designs):
1) IFDMA/Multi-IFDMA transceivers are significantly less complex than LFDMA transceivers; in addition, IFDMA/Multi-IFDMA has better Peak-to-Average Power Ratio (PAPR) than LFDMA;
2) IFDMA/Multi-IFDMA transceivers and OFDMA transceivers are comparable in complexity; but IFDMA/Multi-IFDMA has significantly better PAPR than OFDMA.
\end{abstract}

\begin{IEEEkeywords}
Cooley-Tukey FFT, SC-FDMA, IFDMA, LFDMA, OFDMA, transceiver design, bit-reversal.
\end{IEEEkeywords}

\section{Introduction}
Orthogonal Frequency Division Multiple Access (OFDMA), Localized FDMA (LFDMA), and Interleaved FDMA (IFDMA) are a class of signal modulation and multiple-access techniques. For these techniques, information of users are multiplexed and carried on subcarriers within a shared spectrum \cite{MAarticle0,MAbook,MAarticle}.

Among these schemes, IFDMA has attracted great interest because of its excellent peak-to-average power ratio (PAPR) and its ability to achieve full frequency diversity \cite{IFDMA1,IFDMA2}. Other advantages of IFDMA over OFDMA include smaller sensitivity to frequency offset, and ability to combat frequency nulls \cite{SCBook}.

IFDMA, however, has rigid requirements on subcarrier allocation. Let us refer to a data stream multiplexed onto a set of subcarriers within the shared spectrum using IFDMA as an IFDMA stream. At any one time, the shared spectrum may be occupied by IFDMA streams of multiple users. IFDMA streams are subject to the following two constraints \cite{MAarticle,IFDMA1}:
\begin{itemize}
\item {Constraint 1}: Subcarrier allocation in IFDMA must be carefully orchestrated so that not only the subcarriers of different IFDMA streams are non-overlapping, but also that the subcarriers of each IFDMA stream are evenly spaced (e.g., subcarriers $\{0, 4, 8, ...\}$ are allocated to one IFDMA stream; subcarriers $\{1, 17, 33, ...\}$ to another IFDMA stream).

\item {Constraint 2}: The number of subcarriers of an IFDMA stream must be a divisor of the number of the overall available subcarriers (e.g., if there are $64$ subcarriers, the number of subcarriers of an IFDMA stream must be one of $\{1, 2, 4, 8, 16, 32, 64\}$; the number of subcarriers of a stream cannot be $3$).
\end{itemize}

These two constraints lead to the impressions that for IFDMA:
1) multiplexing and demultiplexing of data streams by the transceiver are complicated;
2) subcarrier allocation is not only complicated, but also very inflexible.

This paper argues that both impressions are not exactly true. In fact, the imposed rigidity belies a nice mathematical structure that can be exploited to simplify both transceiver design and subcarrier allocation. The foundation of our results can be summarized in one sentence:

\vspace{1.0em}
{\it Multiplexing/demultiplexing and modulation/demodulation of IFDMA data streams can be interpreted as smaller FFTs/IFFTs of different scales embedded in the overall Cooley-Tukey FFT \cite{FFT1} recursive decomposition scheme.}
\vspace{1.0em}

This realization gives rise to the following two IFDMA constructions, one related to transceiver design, one related to resource allocation:
\begin{itemize}
\item Greatly simplified transceiver designs that multiplex/demultiplex and modulate/demodulate multiple IFDMA streams in parallel within a single FFT/IFFT module.

\item A bit-reversal subcarrier allocation scheme whereby, instead of focusing on allocating evenly-spaced subcarriers to an IFDMA stream, the focus can move to allocating contiguous ``bins'' to an IFDMA stream. After such bin allocation, a bit-reversal mapping of the bin indexes automatically yields the evenly-spaced subcarrier indexes (e.g., if there are $64$ subcarriers, an IFDMA stream allocated contiguous bins $\{0, 1, 2, 3\}$ will be mapped to subcarriers $\{0, 32, 16, 48\}$, i.e., evenly-spaced subcarriers $\{0, 16, 32, 48\}$).
\end{itemize}

The bit-reversal subcarrier allocation scheme addresses Constraint 1. The details of the scheme, its variants, and their implications for resource allocation are treated comprehensively in our companion paper~\cite{tech1}. The main focus of this paper is on the transceiver designs. A quick summary of the bit-reversal subcarrier allocation scheme is given in Section \ref{sec:III} to provide the context.

Our major contributions and conclusions are as follows:
\begin{enumerate}
\item We devise a class of unified IFDMA transceivers that is compatible with all communication scenarios, whether uplink or downlink, and whether IFDMA or Multi-IFDMA (see contribution 2 below on Multi-IFDMA). Benchmarked against conventional IFDMA transceivers, the new transceiver designs can reduce complexity by at least a factor of $\log{M}$, where $M$ is the number of subcarriers available in the overall IFDMA spectrum.
\item We put forth a Multi-IFDMA scheme that allows each user to be allocated multiple IFDMA streams to address Constraint 2 (e.g., with $M=64$, a user that wants $5$ subcarriers can be allocated two IFDMA streams, one with $4$ subcarriers, one with $1$ subcarrier). With bit-reversal subcarrier allocation and Multi-IFDMA, IFDMA can achieve the same level of resource allocation flexibility as LFDMA and OFDMA.
\item Compared with LFDMA, IFDMA/Multi-IFDMA has better PAPR performance and significantly simpler transceiver designs. Compared with OFDMA, IFDMA/Multi-IFDMA has significantly better PAPR performance and comparable transceiver complexity.
\end{enumerate}

\vspace{0.8em}
\noindent\textbf{Related Work} -- The reader is referred to \cite{MAarticle0,MAbook,MAarticle,SCBook,MA2} for detailed surveys and tutorials on OFDMA, LFDMA, and IFDMA. We only provide a quick overview here to put things in context.

OFDMA \cite{MAbook} is the multiple-user variant of OFDM in which subsets of orthogonal subcarriers are allocated to different users. At the transmitter, OFDMA directly multiplexes and modulates a block of symbols onto the allocated subcarriers via OFDM modulation through an IDFT. At the receiver, after OFDM demodulation through a DFT, individual users' data are extracted from the allocated subcarriers.

Two well-known disadvantages of OFDMA are high PAPR and high sensitivity to frequency offset. In view of this, the so-called single-carrier FDMA (SC-FDMA) has been proposed to tackle these two problems \cite{MAarticle}.

The signal processing of SC-FDMA is very similar to that of OFDMA, with an added component.
The addition at the transmitter is that, before OFDM modulation, SC-FDMA first transforms the block of symbols to be transmitted to the frequency domain by a small DFT. The DFT-transformed symbols are then mapped to the allocated subcarriers.
What follows is the same as in OFDMA, i.e., OFDM modulation via an IDFT operation.
The additional DFT operation spreads each of the transmitted symbols over the allocated subcarriers, and the resulting signal is more like a single-carrier modulated signal, which has lower PAPR compared with OFDMA signals.
At the receiver side, as in OFDMA, after OFDM demodulation via DFT, the signals on the allocated subcarriers are then extracted.
The additional component that is needed is an IDFT operation on the extracted subcarriers then despreads the data back to the original data.

SC-FDMA has two subcarrier mapping schemes: localized and distributed \cite{MAarticle0,IFDMA1}. For localized subcarrier mapping, the DFT-transformed symbols are mapped to contiguous subcarriers. SC-FDMA with localized subcarrier mapping is commonly referred to as LFDMA. For the distributed subcarrier mapping scheme, the DFT-transformed symbols are mapped to non-contiguous subcarriers. In particular, full frequency diversity can be achieved when the occupied subcarriers are evenly distributed over all subcarriers, in which case we have the IFDMA scheme.

IFDMA is a scheme that combines the advantages of both single-carrier and multi-carrier modulated systems \cite{IFDMA1}.
At the transmitter, each data symbol is spread over the allocated subcarriers which are scattered over the entire spectrum, with each time-domain sample of an IFDMA stream retaining the original data symbol characteristics (e.g., a time-domain sample is not an inter-mixing of the original data; it is still QPSK if the original data is QPSK data). As a result, IFDMA, like single-carrier modulated systems, has excellent PAPR characteristics.
At the same time, an IFDMA receiver can equalize the channel in the frequency domain as multi-carrier modulated systems do, obviating the need for complicated time-domain equalizers. There has also been description of IFDMA from a multi-carrier CDMA perspective: IFDMA is CDMA with the roles of spreading sequence and data sequence interchanged \cite{DSCDMA}.

The conventional IFDMA transceiver designs have been presented in many papers \cite{IFDMA1,MAarticle0,MAarticle,SCBook}. We will review these designs and analyze their complexities in Section \ref{sec:II}. Further, we will show in Sections \ref{sec:IV} and \ref{sec:V} that the aforementioned DFT/IDFT multiplexing/demultiplexing operations for general SC-FDMA are unnecessarily complex when it comes to IFDMA. Transceiver designs customized for IFDMA can be much simpler.

A subcarrier allocation scheme for IFDMA was proposed in \cite{Allocation1,Allocation2}, where a tree structure similar to that constructed for orthogonal variable spreading factor (OVSF) code was used to allocate subcarriers to different users based on their requests. However, a prerequisite of this allocation scheme is that Constraint 2 must be satisfied. That is, the number of subcarriers of each IFDMA stream must be a divisor of the number of the overall available subcarriers. The Multi-IFDMA scheme proposed in this paper, on the other hand, further eliminates the inflexibility brought by Constraint 2.

\section{Review of IFDMA and Complexities of Conventional IFDMA Transceivers}\label{sec:II}
\input{SecII.tex}

\section{Overview of Our Key Results}\label{sec:III}
\input{SecIII.tex}

\section{Receiver Design}\label{sec:IV}
\input{SecIV.tex}

\section{Transmitter Design}\label{sec:V}
\input{SecV.tex}

\section{Conclusion}\label{sec:Conclusion}
Interleaved frequency division multiple access (IFDMA) is a multiplexing and signal modulation technique for multi-user systems. It has attracted great interest because of its excellent peak-to-average power ratio (PAPR) compared with orthogonal FDMA (OFDMA) and localized FDMA (LFDMA).
IFDMA, however, has strict constraints on subcarrier mapping:

Constraint 1 -- subcarriers assigned to an IFDMA data stream must be evenly spaced over the spectrum used, and subcarriers assigned to different IFDMA data streams must be orthogonal.

Constraint 2 -- the number of subcarriers assigned to an IFDMA data stream must be a divisor of the total number of available subcarriers.

These constraints led to the impressions that 1) transceiver designs for multiplexing and demultiplexing IFDMA streams must be complex; 2) resource allocation in IFDMA systems must be inflexible.

The results of this paper indicate otherwise.
To debunk 1), we devise new IFDMA transceivers that can reduce complexity by an order of $\log M$ or more. Our designs are inspired by a key finding that multiplexing/demultiplexing and modulation/demodulation of IFDMA data streams are coincident with FFTs/IFFTs of different scales embedded within the overall Cooley-Tukey FFT recursive decomposition.
To debunk 2), we put forth a new Multi-IFDMA scheme that allocates multiple IFDMA streams to each user. In conjunction with a bit-reversal subcarrier allocation algorithm, resource allocation in Multi-IFDMA can be as flexible as those of OFDMA and LFDMA. Importantly, our new transceiver designs are unified designs compatible with both conventional IFDMA and Multi-IFDMA.

Overall, our results yield the following new understandings:
1) IFDMA is amenable to simple transceiver design. In particular, IFDMA transceivers are much simpler than LFDMA transceivers, and comparable to OFDMA transceivers, in complexity.
2) Flexible resource allocation is possible with Multi-IFDMA.
3) Multi-IFDMA, like IFDMA, has better PAPR characteristics than OFDMA and LFDMA.

\appendices
\section{Composite-$M$ IFDMA: Its Relationship with IFDMA Transceiver Design and Resource Allocation}\label{sec:AppA}
\input{AppendixA.tex}

\section{Comparison between multi-IFDMA and PAPR-improved LFDMA and OFDMA}\label{sec:AppB}
\input{AppendixB.tex}

\bibliographystyle{IEEEtran}
\bibliography{References}

\end{document}

%% file: SecII.tex
This section reviews IFDMA to set the context for our work. In particular, we analyze the complexities of conventional IFDMA transceiver designs.

We consider an IFDMA system where multiple end nodes exchange messages with a common access point (AP).
In the uplink (UL), each end node constructs an IFDMA data stream carried over a number of subcarriers and transmits the stream to the AP. The streams of different end nodes are carried on different subcarriers and they arrive at the AP at the same time.
The receiver then extracts the information streams from different end nodes according to their subcarrier allocations. In the downlink (DL), the AP maps different end node's IFDMA data streams to their respective subcarriers and transmits the aggregated IFDMA streams of all end nodes to the end nodes.
The receiver at each end node then extracts its own IFDMA stream.

\begin{figure}[t]
  \centering
  \includegraphics[width=0.8\columnwidth]{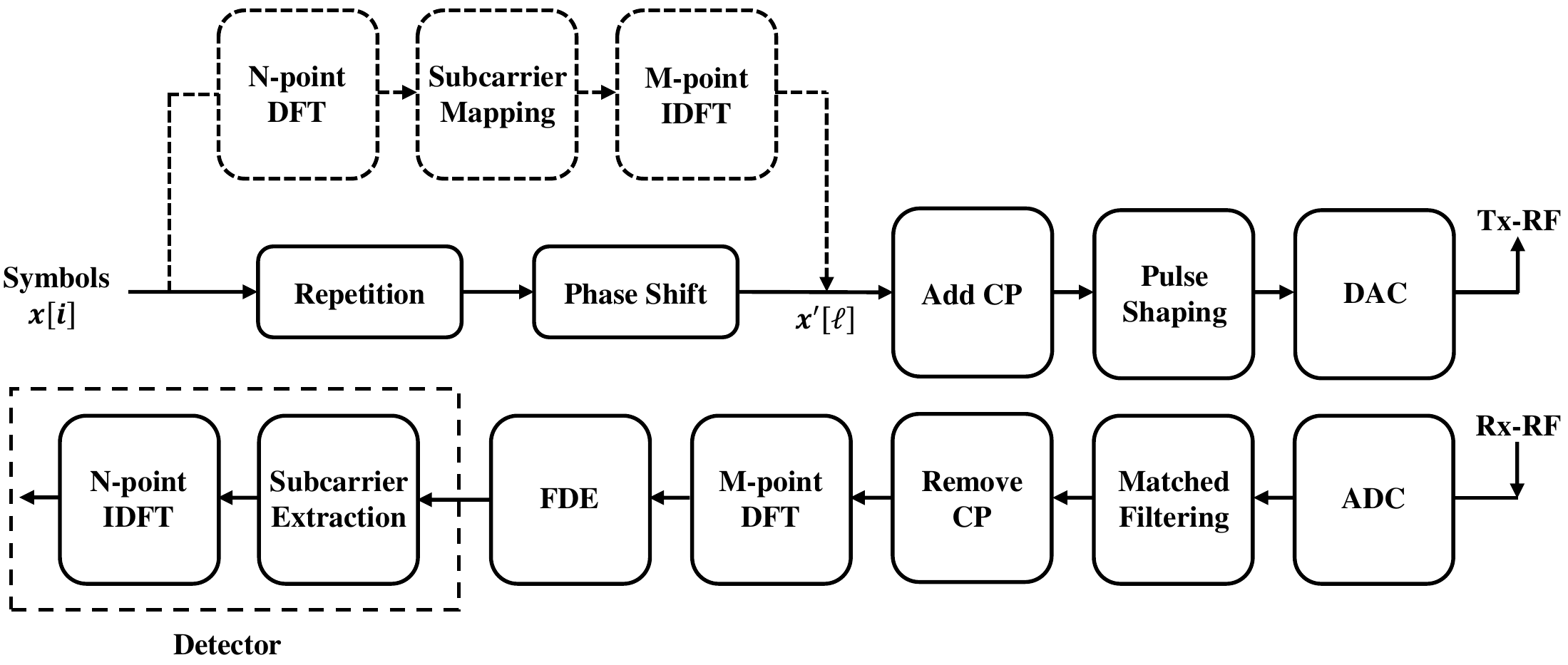}\\
  \caption{The conventional transceiver of a single IFDMA stream \cite{MAarticle0}.}
\label{fig:1}
\end{figure}

Most prior papers present a model of IFDMA as shown in Fig. \ref{fig:1}, which gives the schematic of the transceiver design for a single IFDMA stream.

\subsection{IFDMA: Transmitter}
The transmitter designs for UL and DL in IFDMA systems differ only in the number of streams to be processed. In the UL, the transmitter (end node) has to construct only one IFDMA stream, while in the DL, the transmitter (AP) has to construct multiple IFDMA streams targeted for different end nodes.

Let us first focus on the transmitter design of a single IFDMA stream. Two realizations are given in Fig. \ref{fig:1} \cite{MAarticle0}. One is a frequency-domain realization (the dashed modules), and the other is a time-domain realization (the solid modules).

Let $M$ be the number of subcarriers available in the overall IFDMA system. To construct an IFDMA stream with $N\leq M$ subcarriers, the transmitter (i.e., the end node in UL or the AP in DL) groups the symbols to be transmitted into blocks of size $N$. Denote a block of $N$ time-domain samples by $\{x[n]:n=[N]\}$, where $[N]=0,1,2,...,N-1$. For the frequency-domain realization, the time-domain samples are transformed to frequency-domain samples by an $N$-point DFT, and then mapped to $N$ evenly spaced subcarriers among the $M$ subcarriers. In practice, $M$ can be large and is typically chosen to be a power of $2$ for simple and efficient implementation of radix-$2$ FFT/IFFT. We assume $M$ is a power of $2$ throughout the main body of this paper\footnote{The design principle of this paper applies to any composite number M. Appendix A presents the more general Composite-M IFDMA. Its Relationship with IFDMA Transceiver Design and Resource Allocation are also discussed.}. For an IFDMA stream, the distance between the indexes of two adjacent subcarriers must be $M/N$ \cite{IFDMA1}. That is, $N$ must be a divisor of $M$ so that $M/N$ is an integer\footnote{It is this constraint that gives rise to the impression that IFDMA is very inflexible in terms of resource allocation compared with OFDMA, where any $N$ subcarriers can be allocated to an end node. This concern can be removed with a multi-stream IFDMA scheme elaborated in Section \ref{sec:IIIB}.}. Specifically, the indexes of the $N$ selected subcarriers are
\begin{eqnarray}\label{eq:II1}
\left\{ d+i\frac{M}{N}:i=[N]  \right\},
\end{eqnarray}
where $d\in [M/N]$ is a constant frequency shift.

In this way, the $N$ subcarriers are evenly spaced among the $M$ subcarriers, and the unoccupied subcarriers are filled with $0$. The transmitter then performs an $M$-point IDFT to generate the time-domain IFDMA stream $x'[\ell]:\ell=[M]$. It is easy to verify that\footnote{Most of the nice attributes of IFDMA can be understood from the simple time-domain representation in \eqref{eq:II2}, e.g., the low PAPR property and simple transmitter design when used in UL.} \cite{MAarticle}
\begin{eqnarray}\label{eq:II2}
x'[\ell]=\frac{N}{M}e^{j\frac{2\pi \ell}{M}d}x[\ell~\textup{mod}~N],~~~~\ell=[M].
\end{eqnarray}
When $d=0$, the IFDMA time-domain signal is simply $x'[\ell]=\frac{N}{M}x[\ell~\textup{mod}~N]$, and the subcarriers occupied are $\{iM/N:i=[N]\}$.
In other words, $x'[\ell]$ is simply the block of the original time-domain samples $x[n]$ repeated $M/N$ times normalized by $N/M$.
When $d\neq 0$, the extra phase term $e^{j\frac{2\pi \ell}{M}d}$ in \eqref{eq:II2} corresponds to frequency upshifting of the selected subcarriers.
Eq. \eqref{eq:II2} also indicates that a simpler and more direct time-domain transmitter design is possible. This design bypasses the $N$-point DFT, subcarrier mapping and $M$-point IDFT.

As an example, consider $M = 16$, and let $k=[M]$ be the subcarrier indexes. Suppose an end node A requires $N = 8$ subcarriers. We can set $d = 0$ and allocate subcarriers $\{iM/N:i=[N]\}=\{0,2,4,6,8,10,12,14\}$ to node A. Given a block of $N$ transmitted symbols $\{x_A[n]:n=[8]\}$, the transmitted IFDMA stream is directly generated as
\begin{eqnarray}
x'_A[\ell]=\frac{1}{2} x_A[\ell~\textup{mod}~8],~~~~\ell=[M]. \nonumber
\end{eqnarray}
If another end node B requires $4$ subcarriers, we can then set $d = 1$ and allocate subcarriers $\{1+iM/N:i=[4]\}=\{1,3,5,7\}$ to the IFDMA stream. Given a block of $N$ transmitted symbols $\{x_B[n]:n=[4]\}$, the transmitted IFDMA stream is generated as
\begin{eqnarray}
x'_B[\ell]=\frac{1}{4} e^{j\frac{2\pi \ell}{16}} x_B[\ell~\textup{mod}~4],~~~~\ell=[M]. \nonumber
\end{eqnarray}
As can be seen, the subcarriers of the two IFDMA streams are mutually orthogonal and interleaved.

\noindent\textbf{Complexity Analysis} -- Fig. \ref{fig:1} only shows the generation of one IFDMA stream at the transmitter. In general, if the transmitter has to construct $K$ IFDMA streams (such as that in DL), the blocks between $x[n]$ and $x'[\ell]$ in Fig. \ref{fig:1} have to be repeated $K$ times. Table \ref{tab:1} compares the complexities of the time-domain transmitter and frequency-domain transmitter in terms of the number of complex multipliers required.

\begin{table}[t]
\caption{Complexities of conventional transceivers for IFDMA in terms of number of complex multipliers.}
\center
\setlength{\tabcolsep}{3mm} 
\begin{tabular}{cccc}
\toprule
\rowcolor[HTML]{EFEFEF}
           & {Transmitter (time domian)} & {Transmitter (frequency domain)} & {Receiver} \\
\midrule
\textbf{UL} & $M$    & $\frac{3}{2}M\log_2 M$                        & $\frac{1}{4}M\log^2_2 M+\frac{1}{2}M\log_2 M$                 \\
\textbf{DL} & $M^2$  & $\frac{1}{4}M\log^2_2 M+\frac{1}{2}M\log_2 M$ & $\frac{3}{2}M\log_2 M$                   \\
\bottomrule
\end{tabular}
\label{tab:1}
\end{table}

For UL, we assume that an end node transmits only one IFDMA stream. When implemented in hardware (e.g., FPGA), a time-domain transmitter following \eqref{eq:II2} has to deploy $M$ complex multipliers. A frequency-domain transmitter, on the other hand, needs to deploy many $N$-point DFTs of various sizes, i.e., $\{N = 2^n:n=1,2,...,\log M\}$, so that the end nodes can be allocated different numbers of subcarriers at different times according to their needs. For OFDM modulation, an $M$-point IDFT is required. Let $N=2^n$ and $M=2^m$ where $n$ and $m$ are positive integers. The overall complexity of the frequency-domain implementation is given by
\begin{eqnarray}\label{eq:II3}
\sum_{n=1}^{m}\left(\frac{N}{2}\log_2 N\right)+ \frac{M}{2}\log_2 M \approx \frac{3}{2}M\log_2 M.
\end{eqnarray}
Eq. \eqref{eq:II3} follows from that an $N$-point IDFT requires $\frac{N}{2}\log_2 N$ complex multipliers using Cooley-Tukey FFT \cite{FFT1}.

For DL, the signal transmitted by the AP is an aggregate of multiple IFDMA streams to the end nodes. The number of IFDMA streams equals the number of end nodes. In the extreme case, there are $M$ end nodes, each of which occupies one subcarrier. A time-domain transmitter has to deploy $M^2$ complex multipliers, because the transmitter has to construct $M$ IFDMA streams in the extreme case, each requiring $M$ complex multipliers. The complexity of a frequency-domain transmitter, on the other hand, is
\begin{eqnarray}\label{eq:II4}
\sum_{n=1}^{m}\frac{M}{N}\left(\frac{N}{2}\log_2 N\right) + \frac{M}{2}\log_2 M \approx \frac{1}{4}M\log^2_2 M + \frac{M}{2}\log_2 M,
\end{eqnarray}
because 1) the $M$-point IDFT is common to all IFDMA streams; 2) we need $N$-point IFFTs of various sizes (from $n = 1$ to $n = m$) to accommodate IFDMA steams of different sizes; 3) for a given $N$, up to $M/N$ end nodes can simultaneously request for $N$ subcarriers. Thus, we need $M/N$ $N$-point IDFTs for each possible $N$.

\subsection{IFDMA: Receiver}
For a single IFDMA stream, the receiver can be designed as an exact inverse mapping of the transmitter design in the frequency domain, as shown in Fig. \ref{fig:1}. Specifically, for both UL and DL, the received signal is an aggregate of multiple IFDMA streams, and the $M$-point DFT and frequency-domain equalization (FDE) module are common to all received streams. To extract and decode a single IFDMA stream with $N$ subcarriers, a detector containing a subcarrier extraction module and an $N$-point IDFT is used.

\noindent\textbf{Complexity analysis} -- For DL, there is only one desired IFDMA stream at the receiver (end node). The receiver, however, needs detectors of all possible values of $\{N=2^n:n=1,2,...,m\}$ to decode since the receiver may be allocated IFDMA stream of different sizes at different times. The complexity of the DL receiver is the same as \eqref{eq:II3}.

For UL, the receiver (AP) has to decode all incoming IFDMA streams transmitted by different end nodes. If there are $K$ end nodes, we need $K$ detectors after the FDE, one for each IFDMA stream/end node. When implemented in hardware (e.g., FPGA), the IFDMA receiver not only needs to deploy $N$-point IFFTs of various sizes, but also for each $N$, $M/N$ $N$-point IDFTs are required because up to $M/N$ end nodes can simultaneously request for $N$ subcarriers. The overall complexity of the UL receiver is then the same as \eqref{eq:II4}.

%% file: SecIII.tex
This section discusses subcarrier allocation in IFDMA systems and proposes a new multi-IFDMA system for flexible subcarrier allocation. After that, we give an overview of the complexities of our new IFDMA transceiver designs.
\subsection{Subcarrier Allocation}
A key problem in IFDMA systems is how to allocate subcarriers to end nodes. Compared with OFDMA/LFDMA, IFDMA have stricter requirements on the subcarrier mapping: 1) the subcarriers of an IFDMA stream must be evenly spaced; 2) each IFDMA stream can only have a power-of-$2$ of subcarriers when $M$ is a power of $2$. These two restrictions lead to a misconception that IFDMA is less flexible and efficient than LFDMA.

Our companion paper \cite{tech1} shows that when numbers of subcarriers requested by all end nodes are powers of $2$, subcarrier allocation in IFDMA is no less efficient than in LFDMA and that full loading is possible. The main result is embodied in Theorem \ref{prop:1} below, which states that, provided the sum of the numbers of subcarriers requested by all end nodes do not exceed $M$, there exists a subcarrier allocation scheme that can satisfy the IFDMA evenly-spaced-subcarrier constraint. The proof can be found in our companion paper. Here, we only focus on the implication of Theorem \ref{prop:1}.

\begin{thm}[full loading is possible despite evenly-spaced-subcarrier constraint \cite{tech1}]\label{prop:1}
Consider an IFDMA system with $M=2^m$ subcarriers. Nodes can only request for a power-of-$2$ of subcarriers. Denote by $a_n$ the number of nodes requesting for $2^n$ subcarriers. If the following inequality is satisfied, then we can use a bit-reversal subcarrier allocation algorithm (specified below) to assign subcarriers to nodes such that the IFDMA evenly-spaced-subcarrier constraints are satisfied, i.e., the spacing between the subcarriers assigned to an IFDMA stream with request size $2^n$ must be $2^{m-n}$ -- see \eqref{eq:II1}.
\begin{eqnarray}\label{eq:III1}
\sum_{n=0}^m a_n 2^n \leq 2^m. \nonumber
\end{eqnarray}
\end{thm}
\noindent\textbf{Bit-reversal subcarrier allocation \cite{tech1}}:

1) Instead of allocating subcarriers to nodes, we first focus on allocating $M=2^m$ bins to nodes. Sort all requests in the descending order of their $n$, and allocate bins to them in the order of bin $0$ to bin $M$. That is, the nodes requesting for more subcarriers will be considered first.

2) Perform bit-reversal mapping from bins to subcarriers. If a bin with binary index $b_{m-1}\allowbreak b_{m-2}\allowbreak...\allowbreak b_1b_0$ is allocated to a node, then the subcarrier with binary index $b_0b_1\allowbreak b_{m-2}b_{m-1}$ will be allocated to this node.

For example, suppose that $M = 8$ ($m = 3$), and there are three nodes A, B and C requesting for $2$, $1$, and $4$ subcarriers, respectively. First, we confirm that inequality in Theorem \ref{prop:1} is satisfied, so the bin allocation method is feasible. Second, we sort the nodes according to their request size and allocate bins to them. Node C gets bins $\{0, 1, 2, 3\}$ and node A gets bins $\{4, 5\}$, and node B gets a bin $\{6\}$. Third, we perform bit-reversal mapping to find the subcarriers allocated to the nodes. As shown in Table. \ref{tab:2}, node C gets subcarriers $\{0,2,4,6\}$; node A gets subcarriers $\{1,5\}$, and node B gets subcarriers $\{3\}$.

\begin{table}[t]
\caption{An example of the bit-reversal subcarrier allocation.}
\center
\begin{tabular}{cccc}
\toprule
\rowcolor[HTML]{EFEFEF}
\textbf{Bins}          & \textbf{Binary index of bins} & \textbf{Bit-reversal} & \textbf{Subcarriers} \\
\midrule
0 & $000$ & $000$ & $0$ \\
1 & $001$ & $100$ & $4$ \\
2 & $010$ & $010$ & $2$ \\
3 & $011$ & $110$ & $6$ \\
4 & $100$ & $001$ & $1$ \\
5 & $101$ & $101$ & $5$ \\
6 & $110$ & $011$ & $3$ \\
7 & $111$ & $111$ & $7$ \\
\bottomrule
\end{tabular}
\label{tab:2}
\end{table}

\subsection{Multi-IFDMA}\label{sec:IIIB}
Theorem \ref{prop:1} applies only when the numbers of subcarriers requested by all nodes are restricted to powers of $2$. When the request sizes are non-power-of-$2$, IFDMA does not work since $M/N$ is not an integer, and the constraint that the subcarrier must be evenly spaced cannot be met.

To overcome the problem, we propose to allow each node to be allocated multiple IFDMA streams. For example, consider a node that requires $6$ subcarriers. Obviously, we cannot construct a single IFDMA stream because $6$ is not a power of $2$. However, we can allocate two IFDMA streams to the node, one with $4$ subcarriers and one with $2$ subcarriers. The signal of the node is the sum of the two IFDMA streams.

With the new Multi-IFDMA scheme, the resource allocation of IFDMA system is as flexible as LFDMA/OFDMA (i.e., full loading is possible by Theorem \ref{prop:1}; non-power-of $2$ subcarrier allocation is possible with Multi-IFDMA). We will refer to the conventional IFDMA system that allocates only one IFDMA stream to each end node as the {\it Single-IFDMA}. Although Multi-IFDMA is more flexible than Single-IFDMA, the implementation of Multi-IFDMA is also more complex than the Single-IFDMA if we use the framework of the conventional transceiver in Fig.~\ref{fig:1}.

\begin{table}[t]
\caption{Complexities of conventional transceivers for Multi-IFDMA in terms of number of complex multipliers.}
\center
\setlength{\tabcolsep}{3mm} 
\begin{tabular}{cccc}
\toprule
\rowcolor[HTML]{EFEFEF}
           & {Transmitter (time domain)} & {Transmitter (frequency domain)} & {Receiver} \\
\midrule
\textbf{UL} &                        &                      &                   \\
\textbf{DL} & \multirow{-2}{*}{$M^2$}  & \multirow{-2}{*}{$\frac{1}{4}M\log^2_2 M+\frac{1}{2}M\log_2 M$}  &  \multirow{-2}{*}{$\frac{1}{4}M\log^2_2 M+\frac{1}{2}M\log_2 M$}      \\
\bottomrule
\end{tabular}
\label{tab:3}
\end{table}

Table \ref{tab:3} summarizes the complexities of the multi-IFDMA systems if we use the conventional transceiver designs. Multi-IFDMA is more complex than Single-IFDMA because in Multi-IFDMA, both the transmitter and receiver must be able to process multiple IFDMA streams, whether in UL or DL.

\subsection{Transceiver Design}
The next two sections will present IFDMA transceivers with significantly lower complexity than the conventional designs. Importantly, our transceiver designs are unified designs compatible with both Single-IFDMA and Multi-IFDMA systems.

Our designs are built upon a key finding that multiplexing/demultiplexing of IFDMA streams are equivalent to FFTs of different scales embedded in the Cooley-Tukey FFT decomposition scheme \cite{FFT1}.

\begin{figure}[t]
  \centering
  \includegraphics[width=0.8\columnwidth]{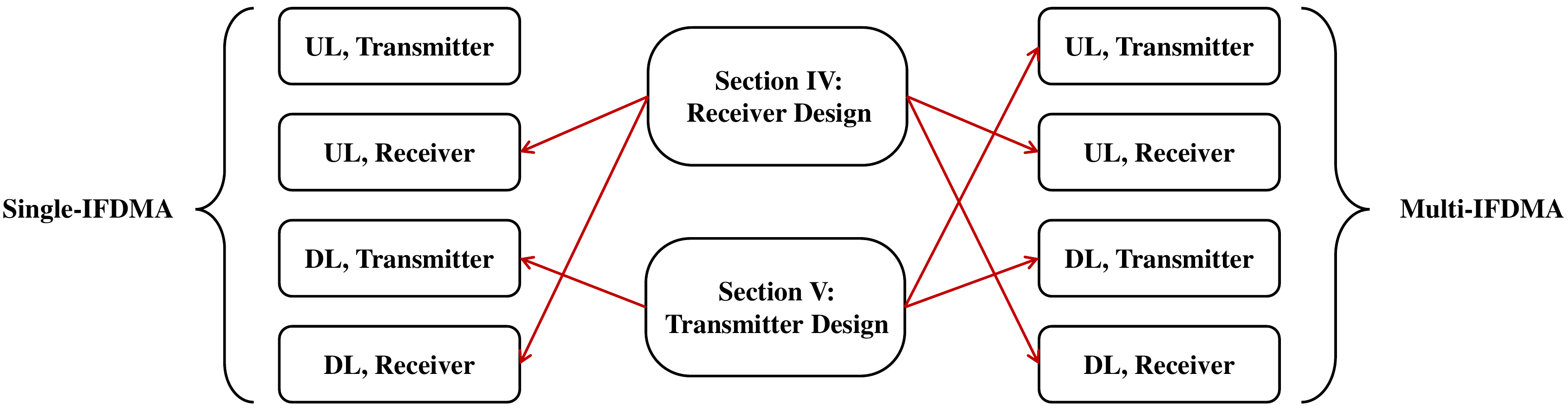}\\
  \caption{Roadmap of subsequent sections.}
\label{fig:2}
\end{figure}

The roadmap of the next two sections of this paper is given in Fig. \ref{fig:2}.
Our transceivers are frequency-domain transceivers. Unlike the conventional frequency-domain design in Fig. \ref{fig:1}, we do not require multiple $N$-point DFTs/IDFTs of various $N$ (the main source of the complexity there). It is possible to use a single $M$-point DFT/IDFT to multiplex (at the transmitter) or demultiplex (at the receiver) the desired IFDMA streams of different sizes, regardless of the number of IFDMA streams and the number of end nodes. The detailed receiver and transmitter designs are presented in Sections \ref{sec:IV} and \ref{sec:V}, respectively.

Tables \ref{tab:4} and \ref{tab:5} summarize the complexities of our transceiver designs and the conventional IFDMA transceiver in terms of the number of complex multipliers required. As shown, our designs require the minimum resources among all designs except for the UL Single-IFDMA case, where an end node only needs to construct a single IFDMA stream. In that case, the time-domain implementation in \eqref{eq:II2} is the simplest way to generate an IFDMA stream. Combing Table \ref{tab:4} and \ref{tab:5}, the complexity of our new transceivers are simpler than the conventional transceiver by at least an order of $\log_2 M$ in all communication scenarios.

\begin{table}[t]
\caption{Comparison of the complexities of different IFDMA receivers.}
\center
\begin{tabular}{cc|c|c|c}
\hline
\rowcolor[HTML]{EFEFEF}
                                                             & \textbf{}   & Conventional Receiver                                           & Our design (with FDE)                                         & Our Design (without FDE)                 \\ \hline
\multicolumn{1}{c|}{}                                        & \textbf{UL} & $\frac{1}{4}M\log^2_2 M+\frac{1}{2}M\log_2 M$                   &                                                               &                                          \\ \cline{2-3}
\multicolumn{1}{c|}{\multirow{-2}{*}{\textbf{Single-IFDMA}}} & \textbf{DL} & $M\log_2 M+\frac{1}{2}M\log_2 M$                                &                                                               &                                          \\ \cline{1-3}
\multicolumn{1}{c|}{}                                        & \textbf{UL} &                                                                 &                                                               &                                          \\ \cline{2-2}
\multicolumn{1}{c|}{\multirow{-2}{*}{\textbf{Multi-IFDMA}}}  & \textbf{DL} & \multirow{-2}{*}{$\frac{1}{4}M\log^2_2 M+\frac{1}{2}M\log_2 M$} & \multirow{-4}{*}{$\frac{1}{2}M\log_2 M+\frac{1}{2}M\log_2 M$} & \multirow{-4}{*}{$\frac{1}{2}M\log_2 M$} \\ \hline
\end{tabular}
\label{tab:4}
\end{table}

\begin{table}[t]
\caption{Comparison of the complexities of different IFDMA transmitters.}
\center
\setlength{\tabcolsep}{3mm} 
\begin{tabular}{cc|c|c|c}
\hline
\rowcolor[HTML]{EFEFEF}
                                                             & \textbf{}   & Transmitter (time domain) & Transmitter (freq. domain)                                       & Our Design                               \\ \hline
\multicolumn{1}{c|}{}                                        & \textbf{UL} & $M$                       & $\frac{3}{2}M\log_2 M$                                           &                                          \\ \cline{2-4}
\multicolumn{1}{c|}{\multirow{-2}{*}{\textbf{Single-IFDMA}}} & \textbf{DL} & $M^2$                     & $\frac{1}{4}M\log^2_2 M +\frac{1}{2}M\log_2 M$                   &                                          \\ \cline{1-4}
\multicolumn{1}{c|}{}                                        & \textbf{UL} &                           &                                                                  &                                          \\ \cline{2-2}
\multicolumn{1}{c|}{\multirow{-2}{*}{\textbf{Multi-IFDMA}}}  & \textbf{DL} & \multirow{-2}{*}{$M^2$}   & \multirow{-2}{*}{$\frac{1}{4}M\log^2_2 M +\frac{1}{2}M\log_2 M$} & \multirow{-4}{*}{$\frac{1}{2}M\log_2 M$} \\ \hline
\end{tabular}
\label{tab:5}
\end{table}

%% file: SecIV.tex
In the conventional design, the receiver for a single IFDMA stream is an exact inverse mapping of the transmitter. An $M$-point DFT followed by a detector are required. When there are $K$ IFDMA streams, the $M$-point DFT is common to all streams, but the detector has to be repeated $K$ times. As given in Table \ref{tab:4}, to accommodate requests of all sizes, the complexity of the receiver is $M\log_2 M+\frac{1}{2}M\log_2 M$ in the DL of the Single-IFDMA systems, and $\frac{1}{4}M\log^2_2 M+\frac{1}{2}M\log_2 M$ otherwise.

In this section, we first dissect the inner structure of IDFT to gain insight into its relationship with IFDMA streams. Based on the insight, we devise a simpler unified detector which uses a single $M$-point IDFT to demultiplex and extract the streams from all transmitters. This unified receiver is compatible with both Single-IFDMA and Multi-IFDMA systems.
\subsection{IDFT Decomposition}
A $2^m$-point DFT/IDFT can be decomposed into two $2^{m-1}$-point DFT/IDFTs, one for odd-indexed samples, and the other for even-indexed samples \cite{FFT1,FFT2}. Specifically, denote by $\Psi_{2^m}$ the $2^m$-point IDFT operation, and $X_{2^m}$ the input vector to the IDFT. We have
\begin{eqnarray}\label{eq:IV1}
\Psi_{2^m}\left(X_{2^m}\right)[\ell]=\frac{1}{2}\Psi_{2^{m-1}}\left(X^e_{2^m}\right)[\ell~\textup{mod}~2^{m-1}]+
\frac{1}{2}W^\ell_{2^m}\Psi_{2^{m-1}}\left(X^o_{2^m}\right)[\ell~\textup{mod}~2^{m-1}],
\end{eqnarray}
where $\ell=[2^m]$, and vectors $X^e_{2^m}$ and $X^o_{2^m}$ contain even-indexed elements and odd-indexed elements of $X_{2^m}$, and $W_{2^m}=e^{j\frac{2\pi}{2^m}}$.

\begin{figure}[t]
  \centering
  \includegraphics[width=0.85\columnwidth]{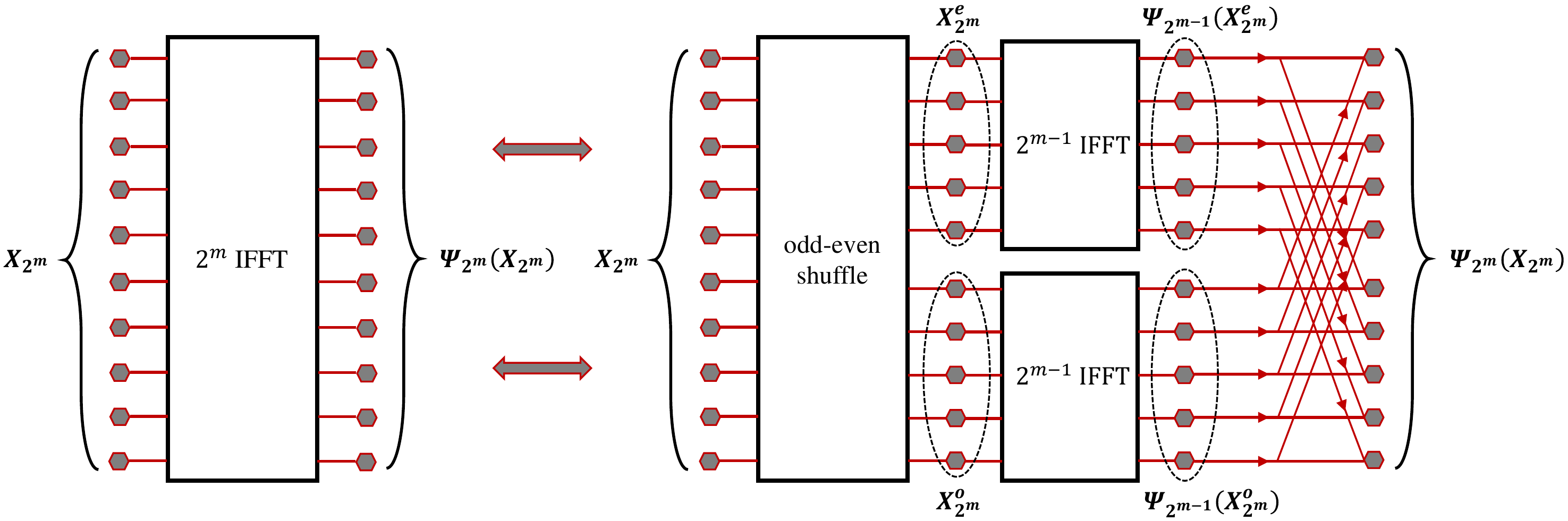}\\
  \caption{Decomposition of $2^m$-point IDFT into two $2^{m-1}$-point IDFTs. The weights in the butterfly networks are omitted.}
\label{fig:3}
\end{figure}

The decomposition in \eqref{eq:IV1} is illustrated in Fig. \ref{fig:3}. As shown, the $2^m$-point IDFT operation on input vector $X_{2^m}$ is equivalent to performing 1) an odd-even shuffle to split $X_{2^m}$ into two vectors $X^e_{2^m}$ and $X^o_{2^m}$; 2) two parallel $2^{m-1}$-point IDFT to get $\Psi_{2^{m-1}}\left(X^e_{2^m}\right)$ and $\Psi_{2^{m-1}}\left(X^o_{2^m}\right)$; and 3) multiple butterfly combinations following \eqref{eq:IV1} to finally get $\Psi_{2^{m}}\left(X_{2^m}\right)$.

In particular, this decomposition process is recursive. Each $2^{m-1}$-point IDFT embedded in the $2^m$-point IDFT can be further decomposed into two $2^{m-2}$-point IDFTs. The overall idea of divide and conquer is the foundation of Cooley-Tukey FFT/IFFT algorithms \cite{FFT1}.

\subsection{IFDMA Detector Design}
Proposition \ref{prop:2} below is a corollary of the above IDFT decomposition. It forms the basis of our IFDMA transceiver designs.

\begin{prop}\label{prop:2}
Denote by $\{X[k]:k=[2^m]\}$ the input to a $2^m$-point IDFT. After $t$ recursive decompositions inside the IDFT, there are $2^t$ parallel $2^{m-t}$-point IDFTs. The inputs to the $d'$-th $2^{m-t}$-point IDFT, $d'=[2^t]$, are $\{X[d+j2^t]:j=[2^{m-t}]\}$, where $d$ is the $t$-bit-reversal mapping of $d'$.
\end{prop}

\begin{NewProof}
Let $b_{m-1}b_{m-2}...b_{1}b_{0}$ be the binary representation of positions occupied by subcarriers at the input of the overall $2^m$-point IDFT. With respect to Fig. \ref{fig:3}, the positions are arranged in an ascending order from top to bottom (i.e., position $00...00$ is at the top and position $11...11$ is at the bottom). Thus, $b_{m-1}b_{m-2}...b_{1}b_{0}$ is also the subcarrier index $k$.

With respect to Fig. \ref{fig:3}, after the first-stage odd-even shuffle of size $2^m$, a subcarrier at original position $b_{m-1}b_{m-2}...b_{1}b_{0}$ will end up at position $b_{0}b_{m-1}b_{m-2}...b_{1}$ (i.e., the odd-even shuffle executes a cyclic right-shift of $b_{m-1}b_{m-2}...b_{1}b_{0}$ so that the new position is $b_{0}b_{m-1}b_{m-2}...b_{1}$). After the two second-stage odd-even shuffles of size $2^{m-1}$ within the $2^{m-1}$-point IDFTs in Fig. \ref{fig:3}, a subcarrier at original position $b_{m-1}b_{m-2}...b_{1}b_{0}$ will end up at position $b_0 b_1 b_{m-1}b_{m-2}...b_{2}$ (i.e., the two smaller odd-even shuffles execute a cyclic right shift of $b_{0}b_{m-1}b_{m-2}...b_{1}$, excluding $b_0$, so that the new position is $b_{0}b_{1}b_{m-1}b_{m-2}...b_{2}$). In general, after the $t$-th odd-even shuffles of size $2^{m-t+1}$, a subcarrier at original position $b_{m-1}b_{m-2}...b_{1}b_{0}$ will end up at position $b_0...b_{t-1}b_{m-1}b_{m-2}...b_{t}$. That is, the position $b_0...b_{t-1}b_{m-1}b_{m-2}...b_{t}$ is the position occupied by subcarrier $b_{m-1}b_{m-2}...b_{1}b_{0}$ when it reaches the input to a $2^{m-t}$-point IDFT.

Consider the $d'$-th $2^{m-t}$-point IDFT. The positions of the inputs to this IDFT are $\{2^{m-t}d'+0,2^{m-t}d'+1,...,2^{m-t}d'+2^{m-t}-1\}$. Let the binary representation of the input positions be $b_0...b_{t-1}b_{m-1}b_{m-2}...b_t$, where $b_0...b_{t-1}$ are fixed to be the binary representation of integer $d'$, and $b_{m-1}b_{m-2}...b_t$ ranges from $00...00$ (corresponding to integer $0$) to $11...11$ (corresponding to integer $2^{m-t}-1$). This binary representation is consistent with the decimal representation.

According to the subcarrier-to-position mapping, the subcarriers occupying the above positions are subcarriers $b_{m-1}b_{m-2}...b_tb_{t-1}...b_0$ where $b_{m-1}b_{m-2}...b_t$ ranges from $0$ to $2^{m-t}-1$, and $b_{t-1}...b_0$ is the $t$-bit-reversal mapping of $b_{0}...b_{t-1}$. Let $d$ be the integer corresponding to $b_{t-1}...b_0$. Thus, the subcarriers mapped to the inputs of the $d'$-th $2^{m-t}$-point IDFT are indexed by $\{d+j2^t:j=[2^{m-t}]\}$.
\end{NewProof}

Consider an IFDMA system with a total of $M=2^m$ subcarriers. Suppose a node A requests for $N$ subcarriers, where $N$ is a power of $2$ (i.e., $N=2^n$, $n = 0, 1, ..., m$). For non-power-of-$2$ $N$, we can use the Multi-IFDMA scheme to divide the $N$ requests to several sub-requests, each of which is a power of $2$.

Following the principle of IFDMA, the subcarriers allocated to node A are indexed by $\{d+j2^{m-n}:j=[2^n]\}$, where constant $d$ can be any integer in $[2^{m-n}]$. To decode this stream at the AP, the detector has to perform 1) demultiplexing, to extract the subcarriers $\{d+j2^{m-n}:j=[2^n]\}$ assigned to node A, and 2) $2^n$-point IDFT, to transform the frequency domain equalized data to the time domain.

Let $m - n = t$ in Proposition \ref{prop:2}. An immediate result is that after $m - n$ recursive decompositions inside the $2^m$-point IDFT, the inputs to the $d'$-th $2^n$-point IDFT are the data on subcarriers $\{d+j2^{m-n}:j=[2^{n}]\}$, $d=[2^{m-n}]$, and $d'$ is a ($m-n$)-bit-reversal mapping of $d$. Notice that these subcarriers are exactly the subcarriers allocated to node A.

The above result implies that the two processes required to extract the data stream from node A, demultiplexing and $2^n$-point IDFT, can be inherently performed within the $2^m$-point IDFT. In other words, a single $2^m$-point IDFT at the receiver can be used to simultaneously demultiplex and decode IFDMA streams of all sizes. This is in contrast to the conventional design implied by Fig. \ref{fig:1}, which uses multiple detectors of different sizes to do so.

Specifically, an IFDMA stream with $2^n$ subcarriers can be extracted after a $2^n$-point IDFT within the overall $2^m$-point IDFT. We can simply insert $2^n$ $1\times 2$ switching elements at the outputs of the $2^n$-point IDFTs to extract the target data on the $2^n$ subcarriers.

\begin{figure}[t]
  \centering
  \includegraphics[width=0.85\columnwidth]{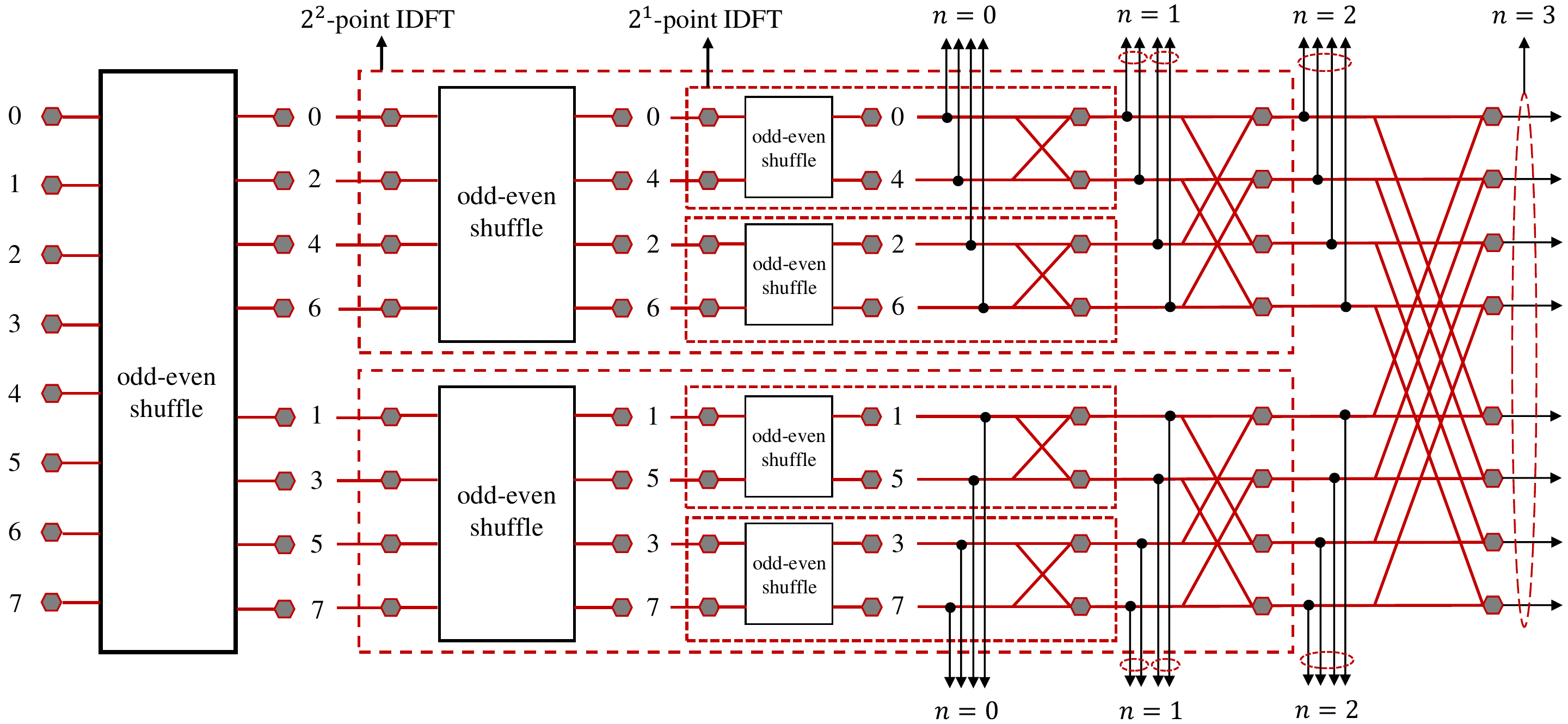}\\
  \caption{Illustration of the IFDMA detector for $M=2^3$. A single M-point IDFT can accommodate IFDMA streams with all possible request size $N=2^n$. Switching elements are inserted within the IDFT network to tap the target streams.}
\label{fig:4}
\end{figure}

Fig. \ref{fig:4}\footnote{We note that the schematic of Fig. \ref{fig:4} also gives an interpretation as to why the bit-reversal subcarrier allocation algorithm works (see Section \ref{sec:II} or \cite{tech1}). In the algorithm, a user is first allocated a set of ``bins'' with contiguous indexes. The bin indexes then undergo a bit-reversal to obtain the corresponding indexes of the interspersed subcarriers allocated to the node. In the schematic of Fig. \ref{fig:4}, the subcarriers allocated are at the input on the left-hand side. The $m$ stages of odd-even shuffles as a whole are equivalent to a bit-reversal mapping. Hence, it maps the subcarriers back to contiguous ``bins''. Throughout the different stages of the subsequent butterfly networks, horizontal line $000$ is bin $000$, horizontal line $001$ is bin $001$, horizontal line $010$ is bin $010$, $...$, and horizontal line $111$ is bin $111$. The butterfly networks do not cause the bin positions to change; just that the content of one bin may be modified by contents of other bins through linear combinations within the butterfly networks. The net effect of the butterfly networks is to convert the frequency-domain signals in the bins to the original time-domain signals. } shows an example of the unified detector where $m = 3$ ($M = 8$). The detector works in the following way:
\begin{enumerate}
\item The data after FDE is indexed from $0$ to $7$ on the LHS of Fig. \ref{fig:4}. The overall structure is a $2^3$-point IDFT, hence an IFDMA stream that occupies $2^3$ subcarriers ($n = 3$) can be extracted at the rear end (if there is such an IFDMA stream allocated to an end node).
\item In the first decomposition, the $2^3$-point IDFT is split into two $2^2$-point IDFTs. In accordance with Proposition \ref{prop:2}, the input to the $d'$-th $2^2$-point IDFT is indexed by $\{d+j2^1:j=[2^2]\}$, where $d$ and $d'$ can be $d = 0$ ($d' = 0$) or $d = 1$ ($d' = 1$). Thus, IFDMA streams that occupy $2^2$ subcarriers ($n = 2$) can be extracted at the outputs of the $2^2$-point IDFTs (if there are such allocated IFDMA streams).
\item In the second decomposition, each $2^2$-point IDFT is again split into two $2^1$-point IDFTs. In accordance with Proposition \ref{prop:2}, the input to the $d'$-th $2^1$-point IDFT is indexed by $\{d+j2^2:j=[2^1]\}$, where $d$ and $d'$ can be $d = 0$ ($d' = 0$), $d = 1$ ($d' = 2$), $d = 2$ ($d' = 1$), or $d = 3$ ($d' = 3$). Thus, IFDMA streams that occupy $2^1$ subcarriers ($n = 1$) can be extracted at the outputs of the $2^1$-point IDFTs (if there are such allocated IFDMA streams).
\item Lastly IFDMA streams that occupy only $1$ subcarrier ($n = 0$) can be extracted right after the odd-even shuffles of size $2$.
\end{enumerate}

\begin{figure}[t]
  \centering
  \includegraphics[width=0.4\columnwidth]{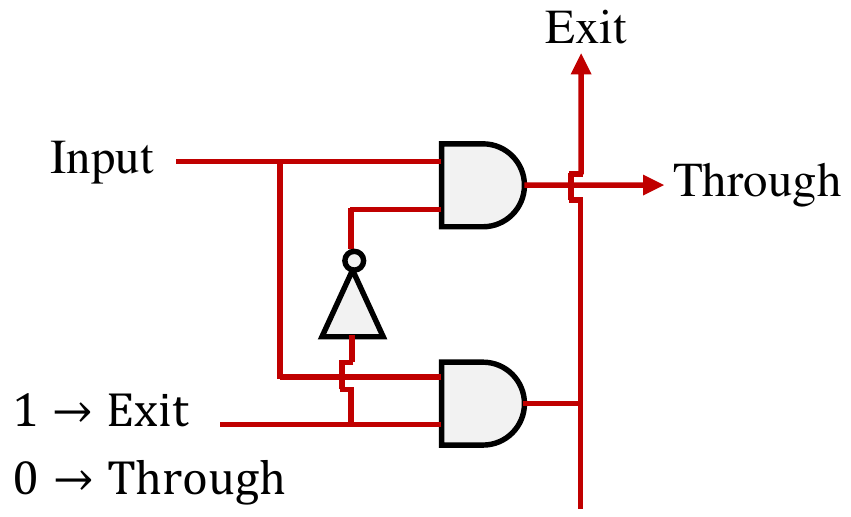}\\
  \caption{A design for the $1\times 2$ switching element using AND and NOT gates \cite{SwitchingBook}. When the control port is set to `0', the ``Input'' will be connected to ``Through''; when the control port is set to `1', the ``Input'' will be connected to `Exit'.}
\label{fig:5}
\end{figure}

Fig. \ref{fig:5} shows a possible design for the $1\times 2$ switching element \cite{SwitchingBook}. As can be seen, the switch element has one input port ``Input'', one control port, and two output ports ``Exit'' and ``Through''. The ``Input'' port is connected to the previous stage within the butterfly networks (see Fig. \ref{fig:4}), and the ``Through'' port is connected to the next stage within the butterfly networks. When the control port is set to `0', the switch is in ``Through'' state, and the input data will flow through the switch to the next stage in the butterfly networks. On the other hand, when the control port is set to `1', the switch is in Exit state, and the input data will be diverted to the ``Exit'' port and be extracted.

Overall, the detector consumes $\frac{1}{2}M\log_2 M$ complex multipliers. Together with the $M$-point DFT, the receiver consumes $M\log_2 M$ complex multipliers (see Table \ref{tab:4} for additional reference). The number of $1\times 2$ switching elements needed to tap the desired streams is $Mlog_2 M$.

\vspace{0.8em}
\noindent\textbf{Selection of taps} -- Another problem after tapping is selection of taps that contain the desired data.
As illustrated in Fig. \ref{fig:4}, there are $M(log_2 M + 1)$ taps in total. Only $M$ of the taps may contain the desired data of the allocated IFDMA streams. Thus, a subsequent switch is required to extract the $M$ target symbols from the $M(log_2 M + 1)$ taps.
We can consider the switching problem separately for each subcarrier.
Specifically, if a particular subcarrier belongs to an IFDMA stream with only one subcarrier, it will be extracted after the $m$-stage odd-even shuffle (before the first butterfly combination); if it belongs to an IFDMA stream with two subcarriers, the particular subcarrier will be extracted after a $2$-point IDFT, etc. Thus, the switching problem for each particular subcarrier is essentially a $(\log_2 M + 1)$ choose $1$ problem. We can then use a $(\log_2 M + 1)\times 1$ switch to export the designed output. For all $M$ subcarriers, a total of $M$ $(\log_2 M + 1)\times 1$ switches are needed.
Rather than considering the design of these switches as a separate problem, we will next present a design in which the tapping and selection of taps are simultaneously performed within the IDFT structure.

\subsection{Design Refinement}
Our first refinement is straightforward. In Fig. \ref{fig:4}, the net effect of the $\log_2 M$ stages of odd-even shuffles of differen sizes is essentially a bit-reversal shuffle. Thus, we can replace all the odd-even shuffles with a single bit-reversal shuffle at the front-end. This is implied by the IDFT decomposition anyway.

The next refinement uses a tapping bus to combine the functions of tapping and selection of taps for each subcarrier. The tapping bus makes use of $2\times 2$ switching elements rather than $1\times 2$ switching elements. A $2\times 2$ switching element connects two input ports ``Input 1'' and ``Input 2'' to two possible output ports ``Through'' and ``Exit''. It is ordinarily set to the ``Through'' state wherein ``Input 1'' is connected to ``Through'', and ``Input 2'' is connected to ``Exit''. In the ``Exit'' state, ``Input 1'' is connected to ``Exit'', and ``Input 2'' is connected to ``Through''. The ``Exit'' state allows a sample to be tapped.

\begin{figure}[t]
  \centering
  \includegraphics[width=0.85\columnwidth]{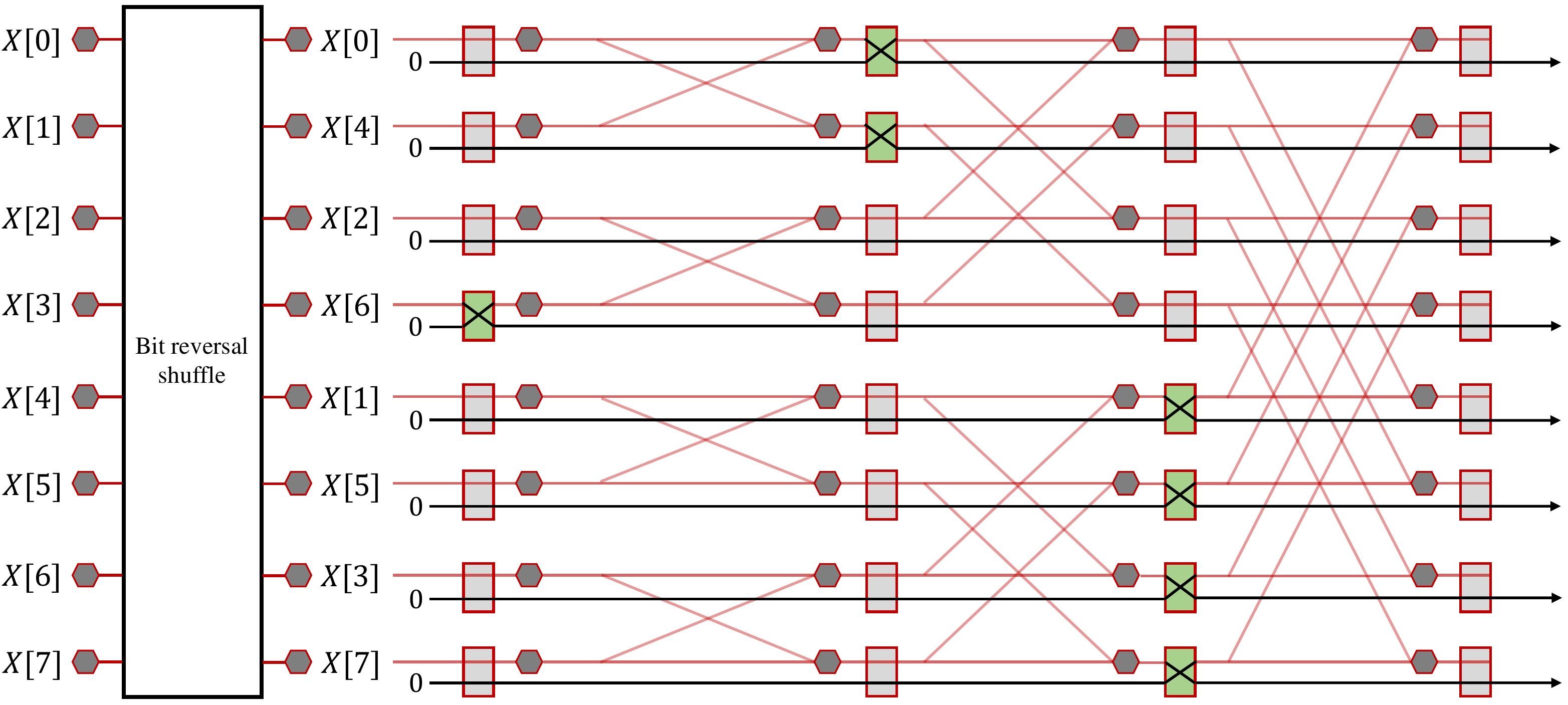}\\
  \caption{An example of the refined $M$-point IDFT detector for $M = 8$. The odd-even shuffles are combined into a single bit-reversal shuffle. The tapping and switching problems are solved together by a tapping-bus design. }
\label{fig:6}
\end{figure}

An example is given in Fig. \ref{fig:6}, where $M = 8$. We assume subcarriers $\{1,3,5,7\}$ are assigned to a node A, subcarriers $\{0,4\}$ are assigned to a node B, subcarrier $\{6\}$ is assigned to a node C.  As shown, after the bit-reversal shuffle, we equip each subcarrier with a tapping bus, into which $(\log_2 M+1)$ $2\times 2$ switching elements are inserted (right after the $2^n$-point IDFT for $n = 0, 1, 2, 3$).

For node A, the data should be extracted at the exit of the $4$-point IDFT. Thus, on the tapping buses of subcarriers $\{1,3,5,7\}$, the $2\times 2$ switching elements right after the $4$-point IDFT are set to the Exit states, while the other switching elements will stay at the Through states. In this way, node A's data can be tapped off at the exit of the tapping bus. Note that the function of tap selection is integrated into the tapping-bus design. The Exit state at a particular stage and the Through state at all other stages basically selects the tap with the Exit state and eliminates the taps with the Through state.

Similarly, for the tapping buses of subcarriers $\{0,4\}$, the $2\times 2$ switching elements right after the $2$-point IDFT are set to the Exit states, while the other switching elements stay in the Through states. Node B's data will be tapped out at the exit of the tapping bus. For the tapping bus of subcarriers $\{6\}$, the first $2\times 2$ switching element is set to the Exit state, the switching elements after stay at the Through states. Note that subcarrier $\{2\}$ is unassigned, hence all the switching elements stay at the Through states.

\begin{figure}[t]
  \centering
  \includegraphics[width=0.65\columnwidth]{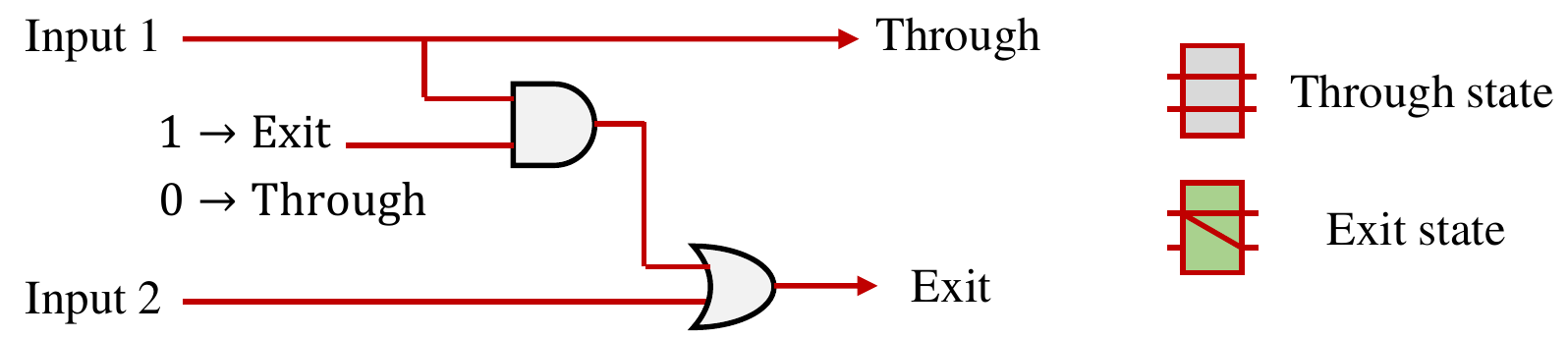}\\
  \caption{A simple design of the $2\times 2$ switching element tailored for the IFDMA detector, using AND and OR gates. When the control port is set to `0', the ``Input 1'' will be connected to ``Through'', and the ``input 2'' will be connected to ``Exit''; when the control port is set to `1', the ``Input 1'' will be connected to both ``Through'' and `Exit' ports.}
\label{fig:7}
\end{figure}

\vspace{0.8em}
\noindent\textbf{Simple Design of $2\times 2$ Switching Element} --
The implementation of a $2\times 2$ switching element is in general more complex than that of a $1\times 2$ switching element in Fig. \ref{fig:4} and Fig. \ref{fig:5}. However, by exploiting an IFDMA property, we can simplify the design of the $2\times 2$ switching elements tailored for our purpose. This custom design comes from an observation of Fig. \ref{fig:6} that if an IFDMA stream being tapped off at a stage is also passed through to the subsequent butterfly combinations at later stages, other IFDMA streams will not be contaminated. This is because the IFDMA streams that could have been affected through the subsequent butterfly combinations must have already been tapped off at the same stage or at a prior stage.
For example, consider subcarrier $X[6]$ in Fig. \ref{fig:6}. Suppose that $X[6]$ gets passed to both outputs Exit and Through. Via the Through output, $X[6]$ will affect the arithmetic at other subcarriers (all $X[i]$ in this example) through the subsequent butterfly combinations. However, the butterfly combinations affected by it are not useful anymore because the associated subcarriers have already been tapped off before they are affected. The reader may ask ``How about $X[2]$?'' If $X[2]$ were associated with an IFDMA stream, given that $X[6]$ is already allocated to another IFDMA stream with one subcarrier, $X[2]$ cannot be associated with an IFDMA stream with more than one subcarrier (this is implied by our bin allocation scheme in Section \ref{sec:II}) and must be an IFDMA stream with one subcarrier, in which case it will be tapped also at the same stage as $X[6]$.

This observation implies that in the Exit state, we can connect ``Input 1'' to both the ``Exit'' and ``Through'' output ports simultaneously. The design of the associated switching element is rather simple, as shown in Fig. \ref{fig:7}. When the control port is set to `0', the switch is in the Through state, ``Input 1'' will be connected to the ``Through'' port, and ``Input 2'' will be connected to the ``Exit'' port. On the other hand, when the control port is set to `1', the switch is in the Exit state, the ``Input 1'' will be connected to both ``Through'' and ``Exit'' ports. We note that in this case, all other $2\times 2$ switching elements of the tapping bus must be set to Through state, so that Input 2 of this $2\times 2$  must contains $0$, and the output of the OR gate will simply follow Input 1.

We can thus replace the $2\times 2$ switching elements in Fig. \ref{fig:6} by our tailored  $2\times 2$ switching elements for our final design. Overall, this IFDMA detector is simply an $M$-point IFFT with tapping buses. It consumes $\frac{1}{2}M\log_2 M$ complex multipliers, and $M(\log_2 M + 1)$ $2\times 2$ switch elements.

\subsection{Extension: Receiver without FDE}
In the original IFDMA receiver design, the input will first go through an $M$-point DFT, followed by a FDE module, and an $M$-point IDFT. Extractions of IFDMA streams are performed within the $M$-point IDFT.

The objective of an equalizer is to compensate for the linear distortion (Inter-Symbol Interference) of the channel. However, after FDE and transformation back to the time domain, the noise on different time samples will be correlated. If we feed such samples to a channel decoder that treats the noise as uncorrelated, we will have suboptimal performance. A more advanced system will use a believe propagation (BP) algorithm or a Viterbi algorithm to perform channel decoding directly on the samples with ISI \cite{BP1,BP2}. For such systems, FDE is not required, and our detector can be further simplified.

We first note that FFT can be designed as a reflection of IFFT, in which case the input data will first go through stages of butterfly networks before a bit reversal shuffle. Specifically, an IFFT can be written in a matrix form as
\begin{eqnarray}\label{eq:IV2}
\bm{\Psi_M(X)}=\bm{S_m S_{m-1} ... S_2 S_1 P_0 X},
\end{eqnarray}
where $\bm{X}$ is the input frequency-domain data to the $M$-point IDFT, $P_0$ is a bit-reversal permutation matrix, and $\{\bm{S_n}:n=1,2,...,m\}$ is the $n$-th stage butterfly network inside the IFFT.

Conversely, an FFT can be written in a matrix form as
\begin{eqnarray}\label{eq:IV3}
\bm{X}=\bm{P^{-1}_0 S^{-1}_1 S^{-1}_2 ... S^{-1}_{m-1} S^{-1}_m \Psi_M(\bm{X})},
\end{eqnarray}
where $\bm{S'_n=S^{-1}_{m+1-n}}$, $n=1,2,...,m$, is the $n$-th stage butterfly network inside the FFT.

In our previous design, the receiver is composed of an $M$-point DFT, an FDE module, and an $M$-point IDFT with tapping buses. If we bypass the FDE, the receiver is an $M$-point DFT followed by an $M$-point IDFT with tapping buses.

Given \eqref{eq:IV2} and \eqref{eq:IV3}, we notice that the rear-end bit-reversal shuffle of the FFT cancels with the front-end bit reversal of IFFT. Operations of some of the stages of their butterfly networks also get canceled. Let us consider an IFDMA stream requesting $2^n$ subcarriers. In our previous design, this stream will be extracted after the $n$-th stage butterfly network inside the $M$-point IDFT. However, it is equivalent to extracting this stream right after the ($m-n$)-th stage butterfly network inside the $M$-point DFT, because all the butterfly combinations of the DFT and IDFT in between cancel out. Essentially, the $M$-point IDFT will not be needed anymore and can be eliminated.

\begin{figure}[t]
  \centering
  \includegraphics[width=0.85\columnwidth]{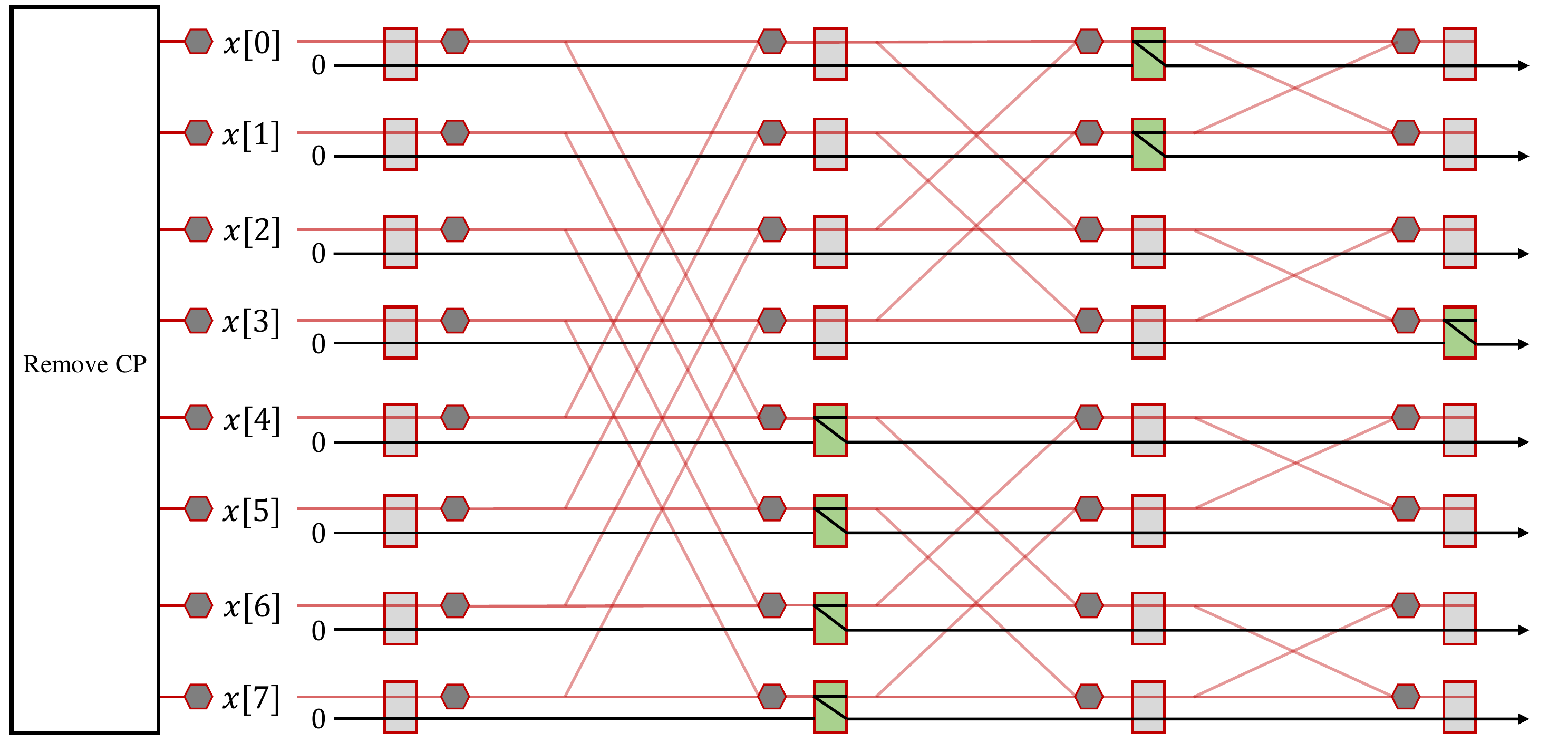}\\
  \caption{Receiver design for systems without frequency domain equalization. Only an $M$-point DFT is required to extract the desired IFDMA streams.}
\label{fig:8}
\end{figure}

As an example, Fig. \ref{fig:8} shows the design corresponding to that of Fig. \ref{fig:6} for which the FDE and $M$-point IDFT are removed. The IFDMA streams are directly extracted within the $M$-point DFT shown in Fig. \ref{fig:8}.

Note that in Fig. \ref{fig:6}, we extract an IFDMA stream with $2^n$ subcarriers after the $n$-th stage butterfly network of the IFFT. In Fig. \ref{fig:8}, we extract the same IFDMA stream at the ($m-n$)-th stage butterfly network of the FFT. Thus, an IFDMA stream with only one subcarrier will have to go through all stages of the FFT, while an IFDMA stream with all $M$ subcarriers does not need to go through the FFT at all.

To conclude this section, let us go back to Table \ref{tab:4} and summarize the advantages of our new receiver design over the conventional receiver design.
\begin{enumerate}
\item Our new receiver design is a unified receiver design. It is compatible with both Single-IFDMA and Multi-IFDMA, and is independent of the number of IFDMA streams and the number of end nodes.
\item Our new receiver design is a much simpler receiver design with a significantly reduced number of complex multipliers. The tapping and switching problems are also solved by the clean tapping-bus design. Only $\frac{1}{2}M(\log_2 M+1)$ $2\times 2$ switching element are needed.
\end{enumerate}

We emphasize that the conventional design also faces a switching problem after the IDFTs of various sizes. That is, the desired streams still need to be selected from among the outputs of the IDFTs of various sizes. The associated switching problem is much more complex than that in our design because of the much higher output dimensionality of the multiple IDFTs.

%% file: SecV.tex
The key insight that led to our simple IFDMA receiver design in the preceding section is that the inner structure of IDFT is coincident with the subcarrier mapping rule of IFDMA. It is not surprising that the same insight can also help simplify IFDMA transmitters. In this section, we first present a simple IFDMA transmitter design. After that, we examine the PAPR characteristics of multi-IFDMA transmitters.

\subsection{Transmitter Design}
Fig. \ref{fig:1} shows two transmitter designs for single-IFDMA, one in the time domain, and one in the frequency domain.
If the transmitter has to construct multiple IFDMA streams (e.g., the DL transmitter or the UL transmitter in Multi-FDMA systems), there are also two ways to generate the aggregated IFDMA signal in the conventional designs. Specifically, the transmitter can
\begin{itemize}
\item generate each data stream separately in the time domain, then sum up the multiple streams thus generated;
\item transform the data of each stream to the frequency domain using an $N$-point DFT (thus, multiple $N$-point DFT of different $N$ will be needed, one for each stream), map the multiple frequency-domain streams to interleaved subcarriers, then transform the aggregated streams back to the time domain using a common $M$-point IDFT.
\end{itemize}

The insight we gained from Section \ref{sec:IV} can be used to simplify the frequency-domain transmitter. In particular, the multiple small $N$-point DFTs before subcarrier mapping can be replaced by a single $M$-point DFT. We have shown in Fig. \ref{fig:4} how the $N$-point IDFTs of different sizes at the receiver can be replaced by an $M$-point IFFT. The same idea applies here. The only difference is that, instead of tapping off the decoded streams, the transmitter has to insert streams into the $M$-point DFT at different stage of the butterfly networks. We could use a DFT design that corresponds to a reflection of the IDFT design in Fig. \ref{fig:4}. With this design, an IFDMA stream with $2^n$ subcarriers should be inserted after the ($m-n$)-th stage butterfly network inside the DFT.

After replacing multiple $N$-point DFTs by one $M$-point DFT, the transmitter is now composed of an $M$-point DFT followed by an $M$-point IDFT. Following \eqref{eq:IV2} and \eqref{eq:IV3}, the mirror-image designs of DFT and IDFT enable us to further simplify the transmitter architecture because the two bit-reversal shuffles (at the rear-end of DFT and the front-end of IDFT) cancel each other and they both can be removed.
Also, the operations of some butterfly networks of DFT and IDFT also cancel each other.
An IFMDA stream with $2^n$ subcarriers, originally be inserted after the ($m-n$)-th stage butterfly network inside the $M$-point DFT, can be inserted right after the $n$-th stage butterfly network inside the $M$-point IDFT. All the butterfly combinations in between cancel out. As a result, the $M$-point DFT can be removed altogether, and we only need an $M$-point IDFT at the transmitter.

\begin{figure}[t]
  \centering
  \includegraphics[width=0.85\columnwidth]{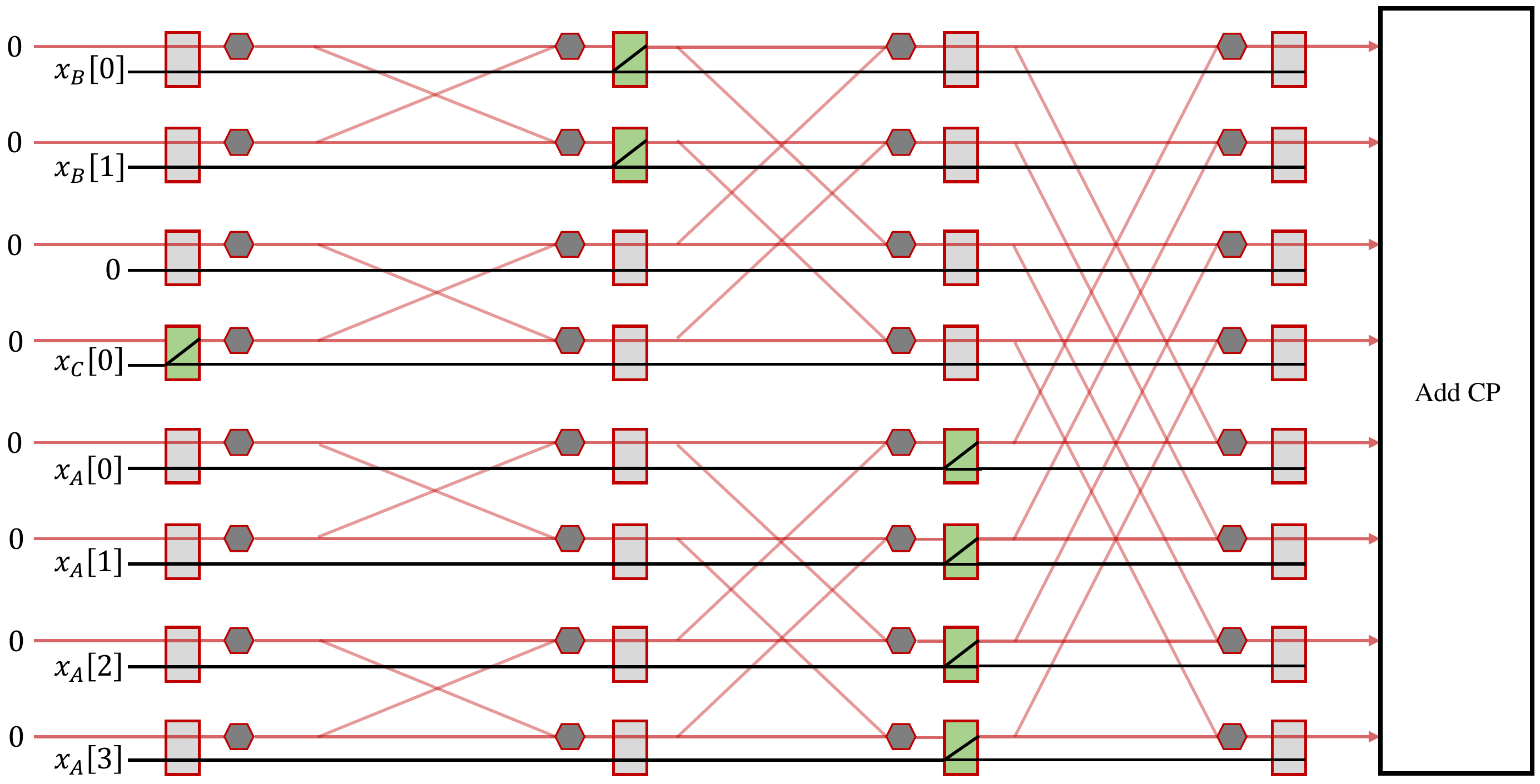}\\
  \caption{An example of the new transmitter design which requires an $M$-point IDFT only.}
\label{fig:9}
\end{figure}

As an example, Fig. \ref{fig:9} presents a transmitter design with $M = 8$ ($m=3$). As shown, the transmitter design requires only a single $M$-point IDFT with tapping buses. The reader may observe that this design is a mirror image of the DFT design of the receiver in Fig. \ref{fig:8}. Over there, the IDFT is redundant and has been removed, here the DFT is redundant and has been removed. In Fig. \ref{fig:9}, the three IFDMA streams are the same as those in Fig. \ref{fig:8}. The first stream occupies $4$ subcarriers indexed by $\{1,3,5,7\}$, the second stream occupies $2$ subcarriers indexed by $\{0,4\}$, and the third stream occupies $1$ subcarrier indexed by $\{6\}$.  In general, an IFDMA stream with $2^n$ subcarriers is inserted right after the n-th stage butterfly network inside the $M$-point IDFT.

Finally, let us go back to Table \ref{tab:5} and compare the complexities of different transmitter designs. As can be seen, our design requires only $\frac{1}{2}M\log_2 M$ complex multipliers at the transmitter. This is the minimal resources required among all other designs except for the UL Single-IFDMA case. In that case, each end node only needs to transmit a single IFDMA stream, and \eqref{eq:II2} gives the simplest implementation.

Fig. \ref{fig:8} and \ref{fig:9} present a final receiver-transmitter pair for IFDMA. An interesting observation is that the input and output to the receiver and transmitter are both time-domain signals. The IFFT/FFT are actually not transforming signals between time domain and frequency domain. For the transmitter, it dispatches each data symbol to a number of multiple time-domain samples, each of which is the original data symbol multiplying by a phase term. For the receiver, it collects the multiple time-domain samples back into the data symbol.
If we combine Table \ref{tab:4} and \ref{tab:5}, the complexity of our new transceivers are simpler than the conventional transceiver by at least an order of $\log M$ in all communication scenarios.

\subsection{PAPR Performance of Multi-IFDMA}
Multi-IFDMA systems achieve the same level of flexibility as OFDMA/LFDMA systems in terms of subcarrier allocation by allowing each end node to be allocated multiple IFDMA streams. The complexity of the Multi-IFDMA transceiver is much lower than that of LFDMA, and is equal to that OFDMA if the IFDMA receiver does not perform FDE.

However, an issue left open is the PAPR property of the Multi-IFDMA systems. In particular, does the Multi-IFDMA transmitter inherit the low PAPR property of the Single-IFDMA transmitter? This subsection addresses this question.

The signal of a single IFDMA stream is a single-carrier modulated signal. With pulse shaping, single-carrier modulated signals in general do not have a Gaussian distribution \cite{PAPR1,PAPR2}. This makes it difficult to analytically derive the statistical properties of a single IFDMA stream. In our case, the transmitted signal in the Multi-IFDMA systems is the sum of multiple (but finite) IFDMA streams, which is also difficult to analyze. Therefore, we resort to Monte Carlo simulations for numerical analysis \cite{MAarticle0}.

In the simulations, we fix $M = 16$, and set different values of $N\leq M$. Our objective is to study the PAPR performance of the Multi-IFDMA system benchmarked against the LFDMA and OFDMA systems. The simulation setups are as follows:
\begin{enumerate}
\item Modulation: QPSK.
\item Subcarrier mapping rule: For OFDMA, the $N$ subcarriers are randomly chosen. For LFDMA, the $N$ contiguous subcarriers are randomly chosen. For Multi-IFDMA, we first partition $N$ to the sum of several integers, each of which is a power of $2$. In particular, the partition is a ``minimal partition'', that is, the partition results in the minimum number of power-of-$2$ integers (e.g., for $N=7$, the partition is $4+2+1$). We construct an IFDMA stream for each power-of-$2$ integer, and sum up the signals of the IFDMA streams in the time domain to construct the Multi-IFDMA stream.
\item Pulse shaping: A root-raised-cosine (RRC) pulse with roll-off factor $\beta=0.5$ is used at the transmitter to shape the transmitted signal. In particular, the RRC pulse is truncated so that it spreads over $20$ symbols only, and it is oversampled $10$ times.
\item PAPR metric: Each packet contains $10$ OFDM symbols, with each OFDM symbol containing $20$ time symbols including CPs. Thus, with the aforementioned $10$-time oversampling, there are $2000$ samples per packet over which we compute the packet's PAPR. We compute the PAPR of 10,000 packets. The performance is measured by the complementary cumulative distribution function (CCDF) of PAPR, as in many other papers \cite{IFDMA1,MAarticle0}. CCDF is defined as the probability that the PAPR is higher than a certain threshold $\eta_0$ (in dB), i.e., $\textup{Pr}(\textup{PAPR}>\eta_0)$.
\end{enumerate}

\begin{figure}[t]
  \centering
  \includegraphics[width=0.7\columnwidth]{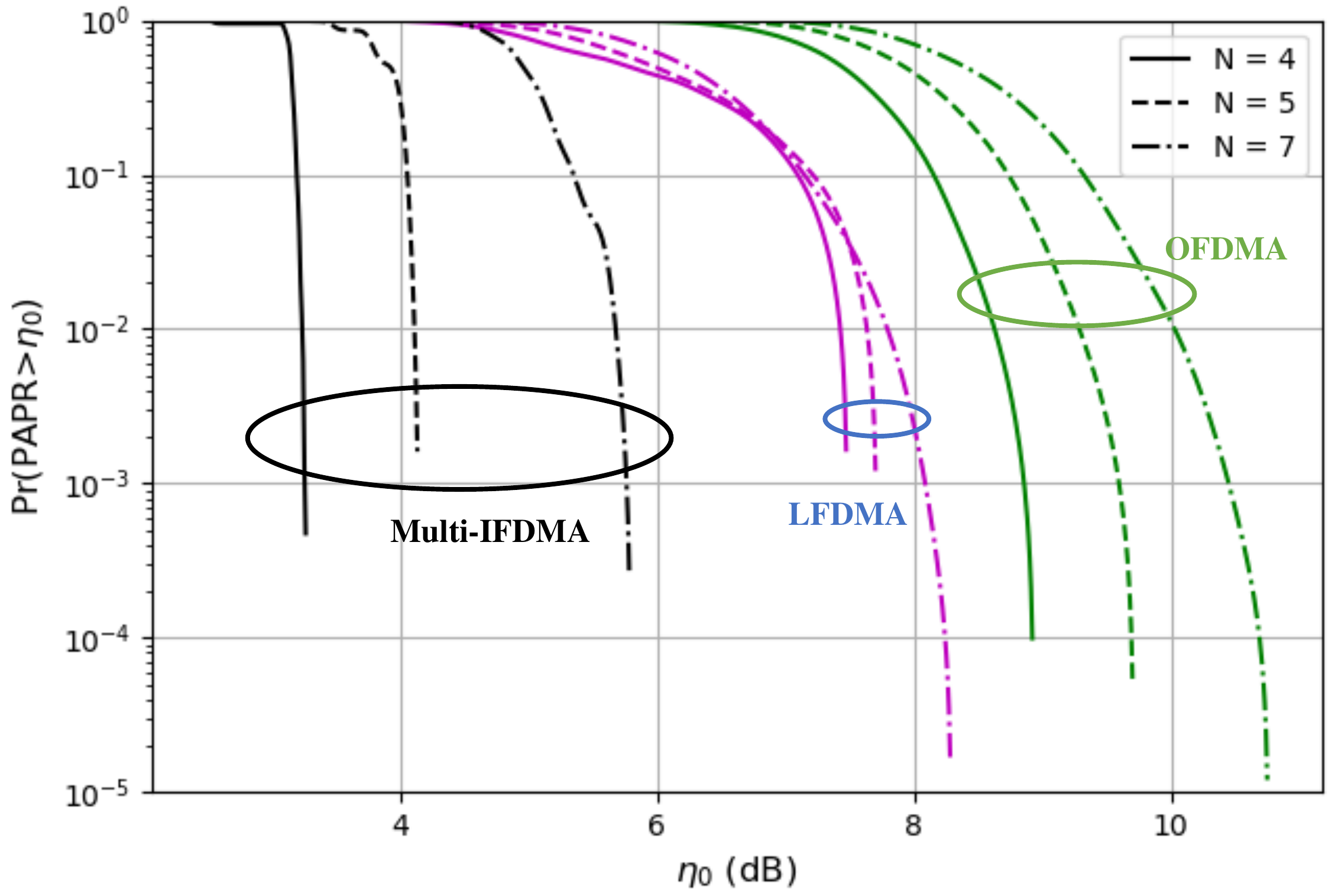}\\
  \caption{The CCDF performance of Multi-IFDMA, LFDMA, OFDMA when $N = 4$, $5$, and $7$.}
\label{fig:10}
\end{figure}

Fig. \ref{fig:10} shows the CCDF performance\footnote{For Multi-IFDMA, the subcarriers are allocated to each stream using our bit-reversal subcarrier allocation scheme. For LFDMA and OFDMA, the CCDF performance is averaged over all possible subcarrier allocation schemes. In the other words, the CCDF performances for LFDMA and OFDMA are the mean performance rather than the worst performance.} of Multi-IFDMA, LFDMA, OFDMA when $N = 4$, $5$, and $7$. When $N = 4$, the Multi-IFDMA stream is a single IFDMA stream. Considering the $99.9$-percentile PAPR $\left(\textup{Pr}(\textup{PAPR}>\eta_0)=10^{-3}\right)$ \cite{MAarticle0}, the CCDF gain of Multi-IFDMA over LFDMA is about $4.2$ dB, and that over OFDMA is about $5.7$ dB. When $N = 5$, the Multi-IFDMA stream consists of two IFDMA streams, one occupying $4$ subcarriers and one occupying $1$ subcarrier. The CCDF performance is $0.9$ dB worse than the $N = 4$ case. This matches with our intuition because the mixing of two IFDMA streams can result in higher amplitude fluctuations. Nevertheless, compared with LFDMA and OFDMA, the $99.9$-percentile PAPR gains of Multi-IFDMA are $3.5$ dB and $5.5$ dB, respectively. Finally, when $N = 7$, the Multi-IFDMA consists of three IFDMA streams, occupying $4$, $2$, and $1$ subcarriers. The CCDF gains over LFDMA and OFDMA are $2.4$ dB and $4.8$ dB, respectively.

\begin{figure}[t]
  \centering
  \includegraphics[width=0.7\columnwidth]{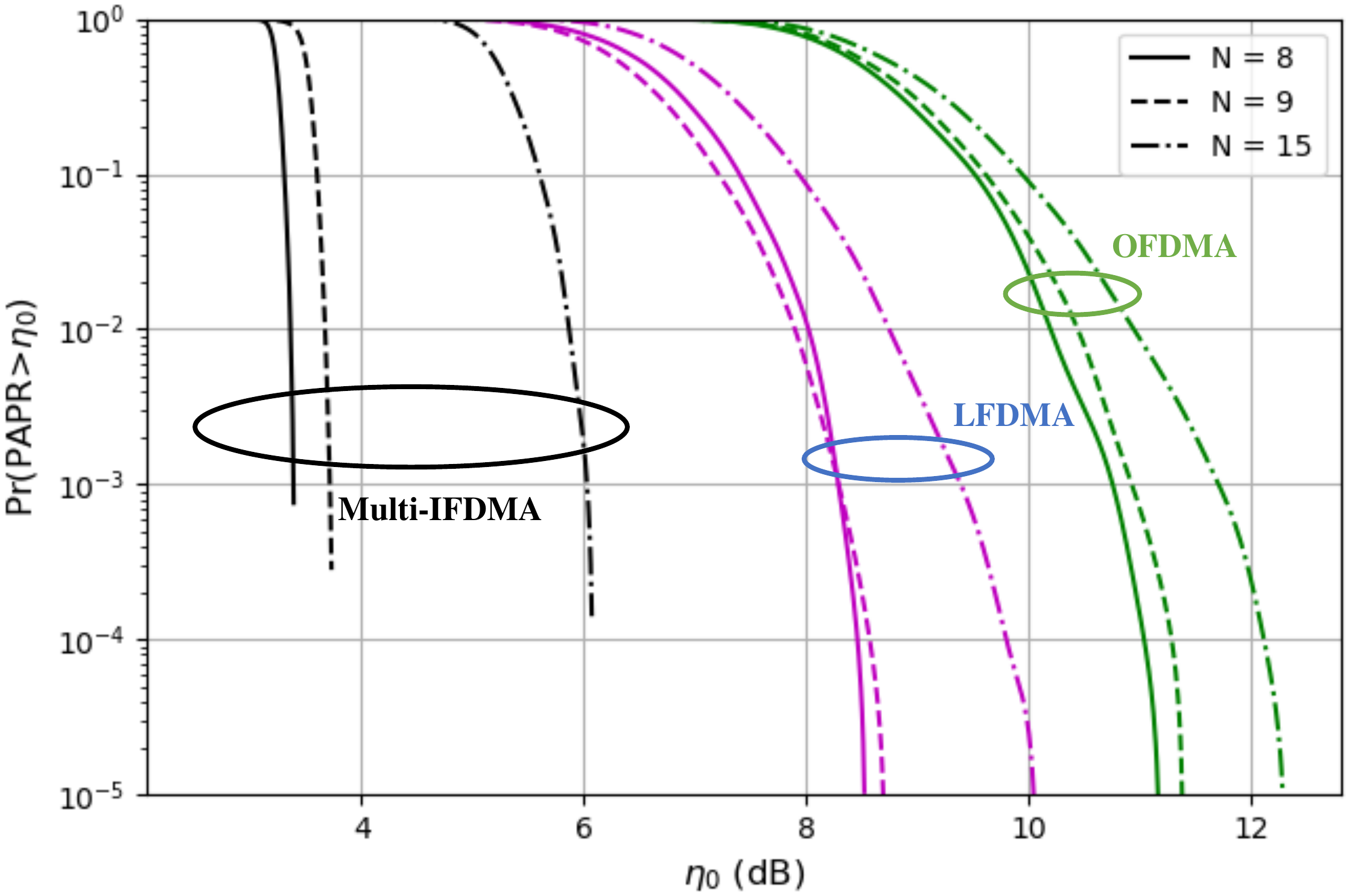}\\
  \caption{The CCDF performance of Multi-IFDMA, LFDMA, OFDMA when $N = 8$, $9$, and $15$.}
\label{fig:11}
\end{figure}

Fig. \ref{fig:11} shows the CCDF performance of Multi-IFDMA, LFDMA, OFDMA when $N = 8$, $9$ and $15$. Similar patterns as in Fig. \ref{fig:10} can be observed. For Multi-IFDMA, a single IFDMA stream has the most favorable PAPR performance when $N$ is a power of $2$. As more and more IFDMA streams are added, the CCDF performance gets worse. The worst PAPR happens when $N=2^n-1$ where $N$ has the most number of partitions. However, even in the worst case, the PAPR performance of Multi-IFDMA is still significantly better than LFDMA or IFDMA. The CCDF gains are up to $3.4$ dB and $5.6$ dB.

Our overall conclusions are as follows. Multi-IFDMA has better PAPR than LFDMA and significantly better PAPR than OFDMA. In terms of transceiver complexity, IFDMA and OFDMA are comparable. Note that for OFDMA, only an IDFT is needed at the transmitter and a DFT is needed at the receiver, and the complexity in terms of number of complex multipliers is therefore $\frac{1}{2}M\log_2 M$. For LFDMA, the complexities of both transmitter and receiver are in the order of $M^2\log_2 M$, if we use the conventional SC-FDMA design (similar to that shown in Fig. \ref{fig:1}).

%% file: AppendixA.tex
In the main body of this paper, we assumed that $M$ is a power of $2$, and showed that the construction/extraction of IFDMA streams of size $2^n$ ($n = 0, 1, 2, ..., \log_2 M$) is embedded in the overall $M$-point Cooley-Tukey FFT/IFFT. In general, this design principle applies to any composite number $M$.

To illustrate, this appendix extends the detector design of Fig. \ref{fig:6} to the general $M$ case. The receiver design for the case without FDE (as in Fig. \ref{fig:8}) and the transmitter design (as in Fig \ref{fig:9}) should be obvious after the presentation below. To conserve space, we omit proofs here and just present the method.

Consider the prime factorization of a composite number $M=\prod_{j=0}^{J-1}p^{q_j}_j$, where $J$ is the number of distinct prime factors in $M$, $p_j$ is a prime factor of $M$, and $q_j$ is the number of times $p_j$ is repeated in $M$. For $M$ that is a power of $2$, there is only one way to perform the FFT/IFFT decomposition. When there is more than one distinct prime factor in $M$, there are more ways to perform the FFT/IFFT decomposition. Specifically, the different orders in which the $M$-point FFT/IFFT is broken down into smaller DFTs/IDFTs yield different decompositions. In general, for $M=\prod_{j=0}^{J-1}p^{q_j}_j$, the number of possible Cooley-Tukey decompositions is given by the multinomial coefficient $\binom{\sum_{j=0}^{J-1}q_j}{q_0,q_1,...,q_{J-1}}$. For example, for $M=12=2^2*3$, there are $\binom{3}{2,1}=3$ possible FFT/IFFT decompositions. The different orders in which decomposition can be performed are $(2,2,3)$, $(2,3,2)$, and $(3,2,2)$.

For an IFDMA system, we can only choose one of the possible FFT decompositions to serve as the architectural bedrock for our transceiver design and subcarrier-allocation algorithm. Intermixing of different FFT decompositions by different users (i.e., different users using different FFT decompositions in their transceiver design) will lead to inconsistencies in the multiplexing and demultiplexing processes, as well as in resource allocation.
For $M=12$, for example, if the order of decomposition is $(2,2,3)$, then the possible sizes for a single IFDMA stream are $12$, $6$, $3$, and $1$; if the order of decomposition is $(2,3,2)$, then the possible sizes are $12$, $6$, $2$, and $1$; and if the order of decomposition is $(3,2,2)$, then the possible sizes are $12$, $4$, $2$, and $1$. The reason is elaborated below.

In the following, let us consider a prime factorization of $M$ of a particular order: $M=\prod_{r=0}^{R-1}f_r$, where $f_r\in\{p_j:j=[J]\}$ and $R=\sum_{j=0}^{J-1}q_j$.

An extension of \eqref{eq:IV1} from power-of-$2$ $M$ to general composite $M$ is as follows:

Given $M=\prod_{r=0}^{R-1}f_r$, an $M$-point IDFT can be decomposed to $f_0$ IDFTs of size $\prod_{r=1}^{R-1}f_r=\frac{M}{f_0}$ by
\begin{eqnarray}\label{eq:AppA1}
\bm{\Psi_M\left(X_M \right)}[\ell] = \frac{1}{f_0}\sum_{d'=0}^{f_0-1}e^{j\frac{2\pi\ell d'}{M}}\bm{\Psi_{M/f_0}\left(X^{d'}_{M/f_0}\right)}[\ell~\textup{mod}~\frac{M}{f_0}] \nonumber
\end{eqnarray}
where $\bm{X_M}$ is the input sequence of length $M$ to the $M$-point IDFT, and $\bm{X^{d'}_{M/f_0}}$ is a subsequence of $\bm{X_M}$ serving as the input sequence to the $d'$-th $M/f_0$-point IDFT, $d'=[f_0]$. The $M/f_0$ elements of $\bm{X^{d'}_{M/f_0}}$ is given by $\bm{X^{d'}_{M/f_0}}[j]=\bm{X_M}[f_0 j+d']$, $j=[M/f_0]$.

\begin{figure}[t]
  \centering
  \includegraphics[width=0.9\columnwidth]{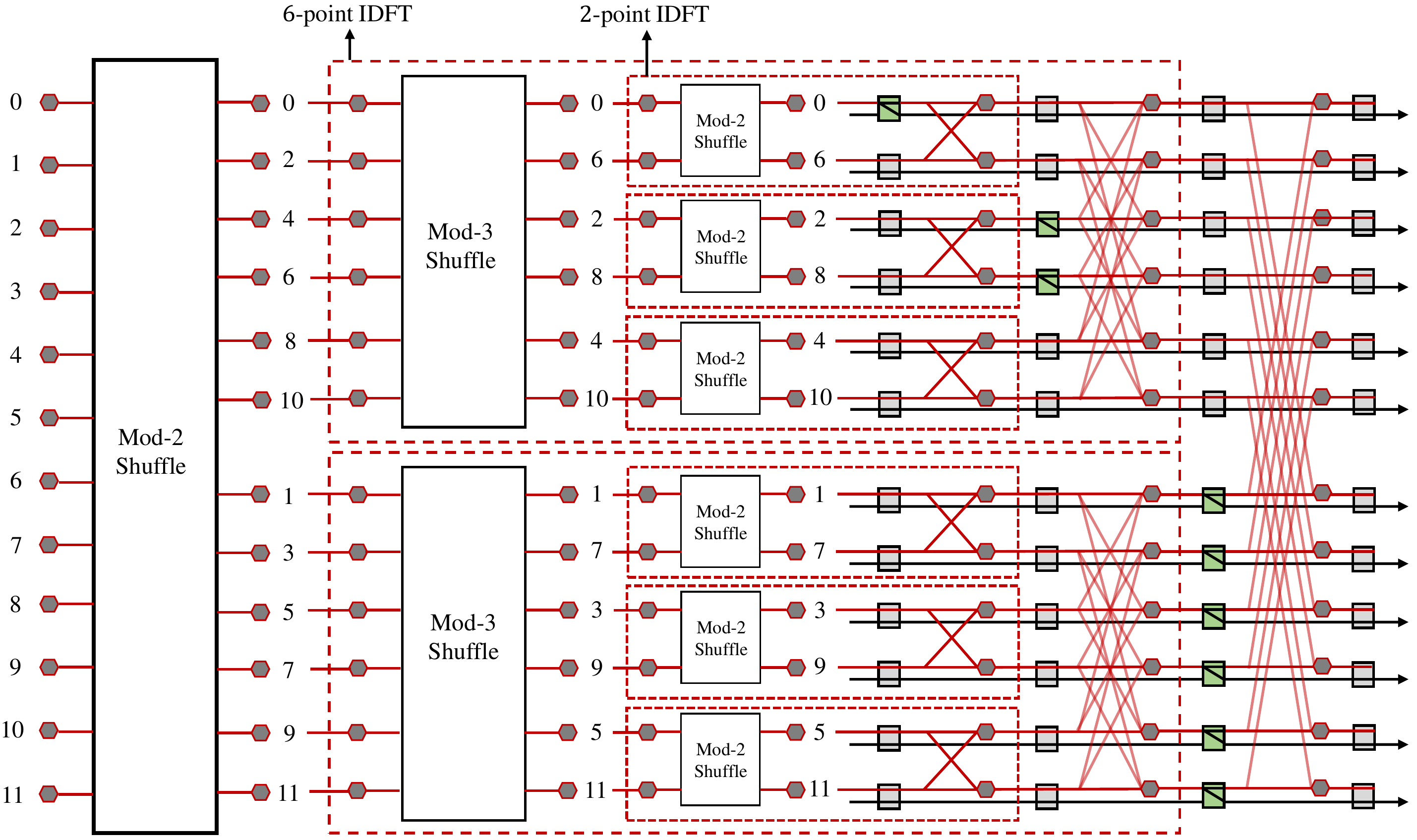}\\
  \caption{A detector design for $M = 12$.}
\label{fig:12}
\end{figure}

The above decomposition is recursive: an $M/f_0$-point IDFT can be further decomposed as $f_1$ IDFTs of size $M/f_0f_1$,  an $M/f_0f_1$-point IDFT can be further decomposed as $f_2$ IDFTs of size $M/f_0f_1f_2$, and so on and so forth.  As in Fig. \ref{fig:6}, the IFDMA streams of different sizes can be extracted after different stages of smaller IDFT inside the M-point IDFT.

An example of the detector design is given in Fig. \ref{fig:12}. In this example, $M = 12$, and the order of decomposition of M is $(2,3,2)$ (i.e., $f_0 = 2$, $f_1 = 3$, $f_2 = 2$). The possible sizes of subcarrier allocations for single-stream IFDMA are $1$, $2$, $6$, and $12$. In the figure, the Mod-$2$ shuffle is just the odd-even shuffle; in general, a Mod-$I$ shuffle is a shuffle in which the inputs are grouped at the output according to $i~\textup{mod}~I$, where $i$ is the input index.

At the transmitter, instead of bit-reversal, a digit-reversal subcarrier allocation scheme can be used to allocate subcarriers to different end nodes (see our companion paper \cite{tech1} for detailed discussions). In the example of Fig. \ref{fig:12}, we assume there are three end nodes. Node A is allocated six subcarriers indexed by $\{1, 3, 5, 7, 9, 11\}$, node B is allocated two subcarriers indexed by $\{2, 8\}$, and node C is allocated one subcarrier indexed by $\{0\}$.

As can be seen from Fig. \ref{fig:12}, node A's IFDMA stream can be extracted after a $6$-point IDFT, node B's IFDMA stream can be extracted after a $2$-point IDFT, and node C's IFDMA stream can be extracted before any butterfly networks.

%% file: AppendixB.tex
People have proposed various PAPR-reduction techniques to reduce the high PAPR of OFDM, among which clipping is the most widely implemented technique thanks to its simplicity.
In this appendix, we perform additional simulations to evaluate the PAPR performance of multi-IFDMA, LFDMA and OFDMA with the clipping technique.

Denote by $x[m]$ the discrete-time signal after pulse shaping. The PAPR of $x[m]$ is given by
\begin{eqnarray}
\rho = \frac{\max_m |x[m]|^2}{\mathbb{E}\left[|x[m]|^2 \right]}.
\end{eqnarray}
The clipped signal is represented as
\begin{eqnarray}
x'[m] = \left\{
\begin{aligned}
x[m], & & \textup{if}~|x[m]|\leq\Gamma \\
\Gamma e^{j\textup{Arg}(x[m])}, & & \textup{if}~|x[m]|>\Gamma,
\end{aligned}
\right.
\end{eqnarray}
That is, we clip $x[m]$ so that the magnitude of the clipped signal does not exceed $\Gamma$ (in this process, the phase of $x[m]$ is preserved). The threshold $\Gamma$ is given by
\begin{eqnarray}
\Gamma = \alpha \sqrt{\mathbb{E}\left[|x[m]|^2 \right]},
\end{eqnarray}
where $\alpha$ is a clipping ratio measuring the severity of clipping. Before clipping, we normalize the power of IFDMA, LFDMA, and OFDMA signal so that for each of the scheme, $\mathbb{E}\left[|x[m]|^2 \right]=1$.

\begin{figure}[t]
  \centering
  \includegraphics[width=0.6\columnwidth]{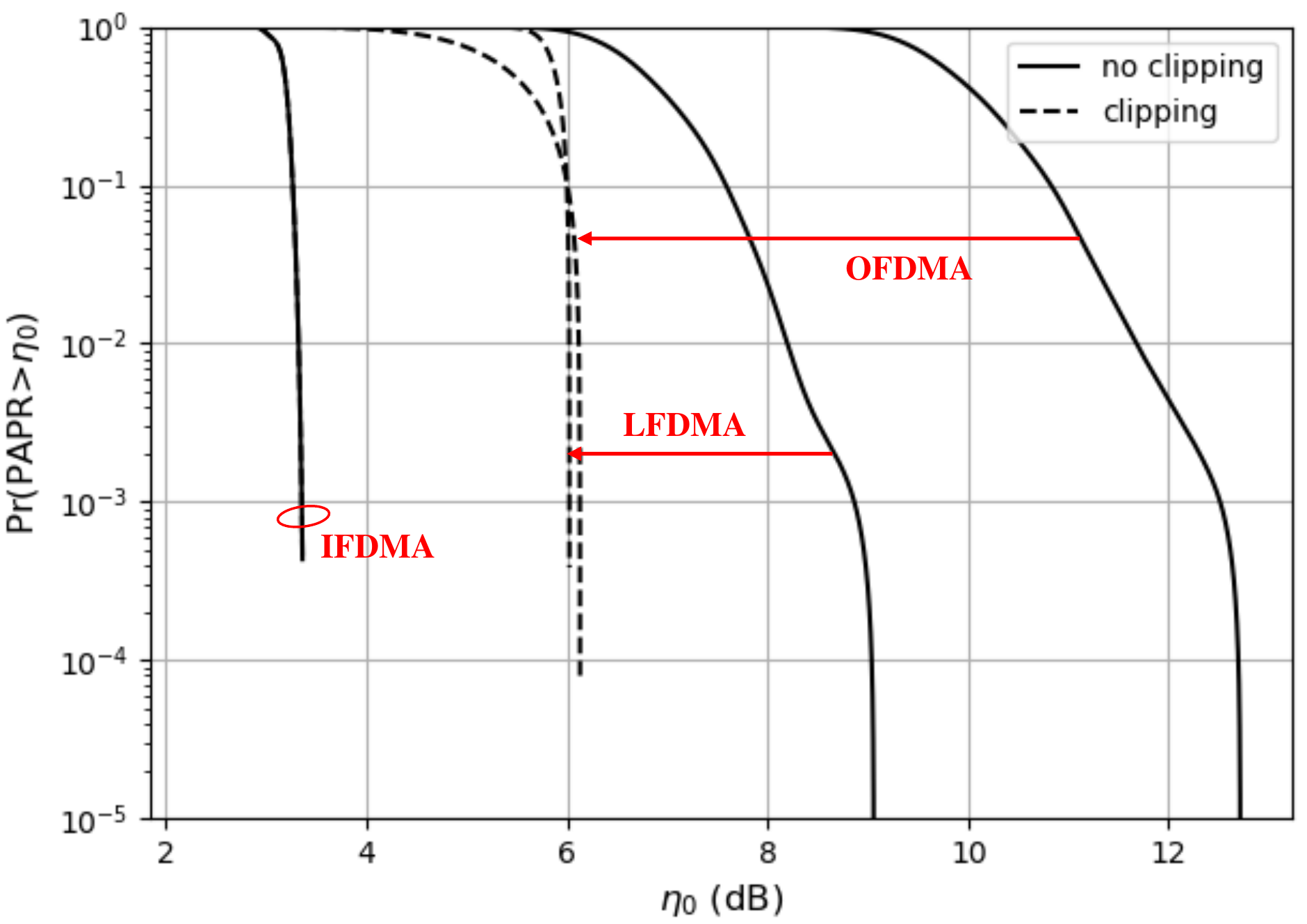}\\
  \caption{CCDF of PAPR with and without clipping, where $N=65$, $M=128$, $\alpha=2$.}
\label{FigR1}
\end{figure}

\begin{figure}[t]
  \centering
  \includegraphics[width=0.6\columnwidth]{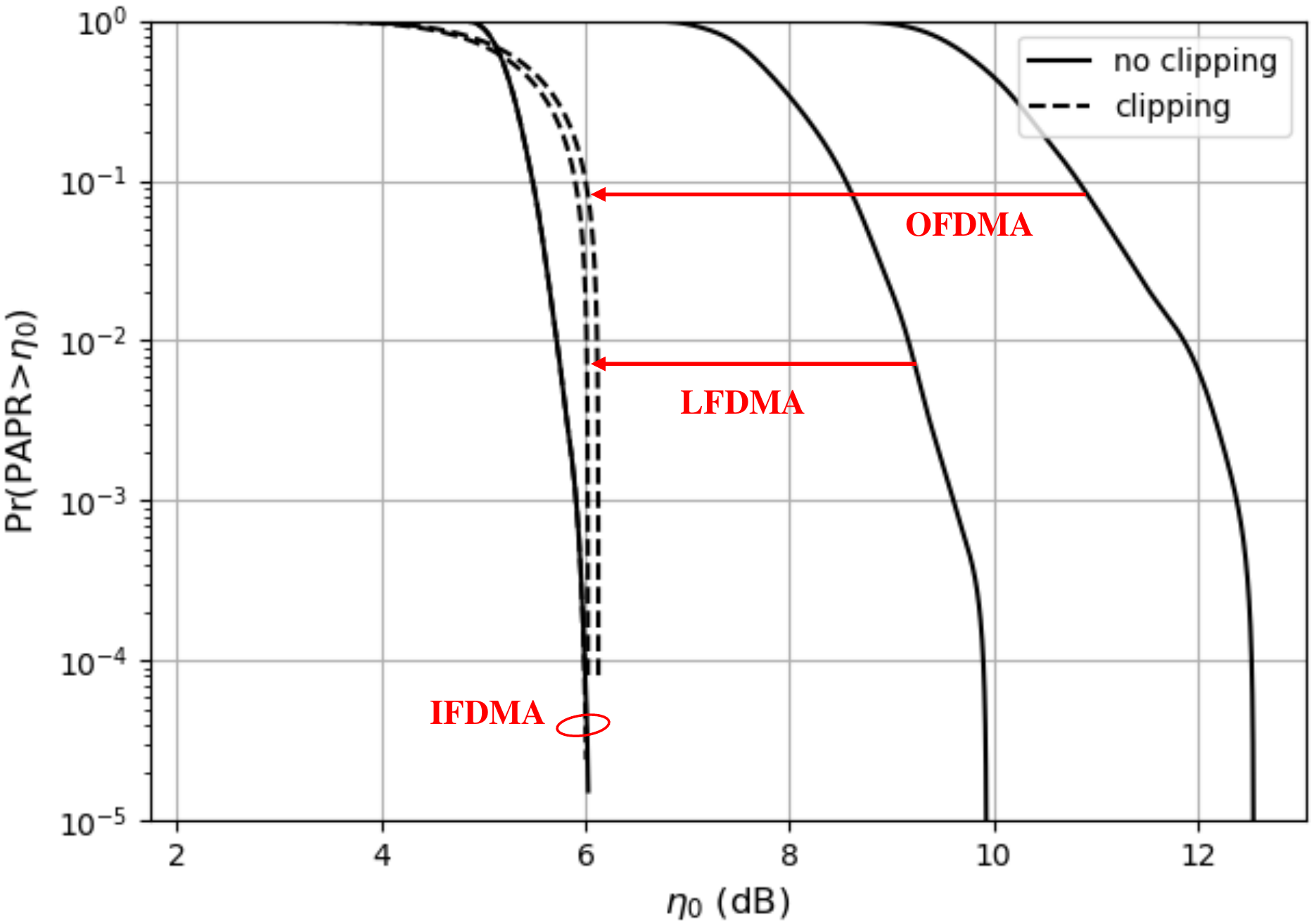}\\
  \caption{CCDF of PAPR with and without clipping, where $N=127$, $M=128$, $\alpha=2$.}
\label{FigR2}
\end{figure}

In Fig. \ref{FigR1} and \ref{FigR2}, we present the simulated PAPR performance of OFDMA, LFDMA and multi-IFDMA with and without clipping. In the simulation, $\alpha=2$, the total number of available subcarriers $M = 128$, and the number of subcarriers requested by users are set to $N = 65$ and $N = 127$, respectively. Note that for multi-IFDMA, $N = 65$ gives us the best PAPR performance because the transmitted signal is a sum of only two IFDMA streams ($65 = 64 + 1$); $N = 127$, on the other hand, gives us the worst PAPR performance because the transmitted signal is a sum of seven IFDMA streams ($127 = 64+32+16+8+4+2+1$), hence the amplitude fluctuates the most.

Considering the $99.9$-percentile PAPR, with clipping, the PAPR performances of LFDMA and OFDMA are improved by around $3$ dB and $7$ dB, respectively. On the other hand, clipping has no effect on multi-IFDMA because the amplitude of the signal is always smaller than the threshold, i.e., $|x[m]|\leq\Gamma=2$.

In the $N = 65$ case, the gap between the multi-IFDMA and the PAPR-improved LFDMA/OFDMA is still large.

We then consider the $N = 127$ case (the worst case for multi-IFDMA). As can be seen from Fig.~\ref{FigR2}, with clipping, the PAPR performance of LFDMA and OFDMA is almost the same as that of multi-IFDMA. However, the improvements do not come without cost. For the clipping technique, the BER performance will suffer.

\begin{figure}[t]
  \centering
  \includegraphics[width=0.6\columnwidth]{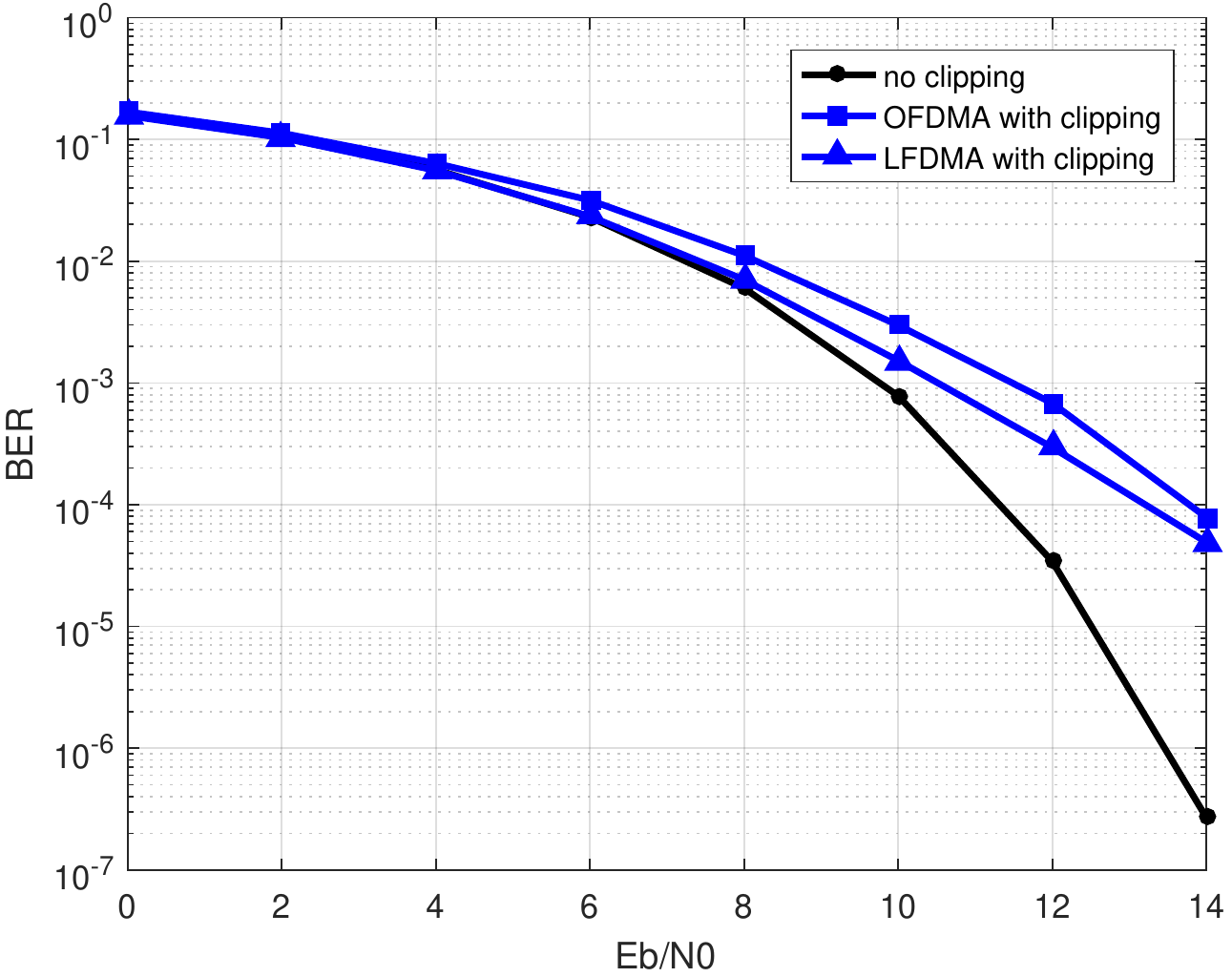}\\
  \caption{BER of multi-IFDMA, LFDMA, and OFDMA systems in AWGN channel with and without clipping, where $N=127$, $M=128$, $\alpha=2$. }
\label{FigR3}
\end{figure}

Fig. \ref{FigR3} compares the BER performance of the above there systems in an AWGN channel with and without clipping. In the simulation, $N$ is set to $127$ and $\alpha$ is set to $2$ so that the PAPR performance of the three systems are almost the same. In particular, to plot the ``no clipping'' curve, we assume ideal amplifier with no nonlinearity, hence the transmitter does not need to clip the amplitude of the signal. This curve corresponds to either multi-IFDMA, LFDMA or OFDMA system without clipping because their performance is the same in this case.

If we clip the signal, however, the BER performance of both LFDMA and OFDMA systems degrades. We do not plot the multi-IFDMA curve in the figure because clipping has no effect on multi-IFDMA (i.e., the multi-IFDMA signal is not clipped at all when $\alpha=2$)  and its BER curve coincides with the ``no clipping'' curve.

In summary, the PAPR performance of LFDMA and OFDMA can be improved by much with clipping, but at the expense of the BER performance. Multi-IFDMA is still preferred compared with PAPR-improved LFDMA and OFDMA. In particular, if we aim for ultra-reliable communication with very low BER, the performance gaps beween Multi-IFDMA and LFDMA/OFDMA will be large. 

%% file: Main.bbl
\begin{thebibliography}{10}
\providecommand{\url}[1]{#1}
\csname url@samestyle\endcsname
\providecommand{\newblock}{\relax}
\providecommand{\bibinfo}[2]{#2}
\providecommand{\BIBentrySTDinterwordspacing}{\spaceskip=0pt\relax}
\providecommand{\BIBentryALTinterwordstretchfactor}{4}
\providecommand{\BIBentryALTinterwordspacing}{\spaceskip=\fontdimen2\font plus
\BIBentryALTinterwordstretchfactor\fontdimen3\font minus
  \fontdimen4\font\relax}
\providecommand{\BIBforeignlanguage}[2]{{%
\expandafter\ifx\csname l@#1\endcsname\relax
\typeout{** WARNING: IEEEtran.bst: No hyphenation pattern has been}%
\typeout{** loaded for the language `#1'. Using the pattern for}%
\typeout{** the default language instead.}%
\else
\language=\csname l@#1\endcsname
\fi
#2}}
\providecommand{\BIBdecl}{\relax}
\BIBdecl

\bibitem{MAarticle0}
H.~G. Myung, ``Introduction to single carrier \textup{FDMA},'' in
  \emph{EURASIP}.\hskip 1em plus 0.5em minus 0.4em\relax IEEE, 2007, pp.
  2144--2148.

\bibitem{MAbook}
H.~Holma and A.~Toskala, \emph{\textup{\it LTE} for \textup{\it UMTS}:
  \textup{\it OFDMA} and \textup{\it SC-FDMA} based radio access}.\hskip 1em
  plus 0.5em minus 0.4em\relax John Wiley \& Sons, 2009.

\bibitem{MAarticle}
H.~G. Myung, J.~Lim, and D.~J. Goodman, ``Single carrier \textup{FDMA} for
  uplink wireless transmission,'' \emph{IEEE Veh. Technol. Mag.}, vol.~1,
  no.~3, pp. 30--38, 2006.

\bibitem{IFDMA1}
T.~Frank, A.~Klein, and E.~Costa, ``\textup{IFDMA}: a scheme combining the
  advantages of \textup{OFDMA} and \textup{CDMA},'' \emph{IEEE Wireless
  Commun.}, vol.~14, no.~3, pp. 9--17, 2007.

\bibitem{IFDMA2}
B.~Lin, X.~Tang, H.~Yang, Z.~Ghassemlooy, S.~Zhang, Y.~Li, and C.~Lin,
  ``Experimental demonstration of \textup{IFDMA} for uplink visible light
  communication,'' \emph{IEEE Photonics Technol. Lett.}, vol.~28, no.~20, pp.
  2218--2220, 2016.

\bibitem{SCBook}
F.~El-Samie, F.~Al-Kamali, A.~Al-Nahari, and M.~Dessouky, \emph{SC-FDMA for
  mobile communications}.\hskip 1em plus 0.5em minus 0.4em\relax CRC press,
  2016.

\bibitem{FFT1}
J.~W. Cooley and J.~W. Tukey, ``An algorithm for the machine calculation of
  complex \textup{F}ourier series,'' \emph{Math. Comput.}, vol.~19, no.~90, pp.
  297--301, 1965.

\bibitem{tech1}
Y.~Shao and S.~C. Liew, ``Flexible resource allocation for interleaved
  frequency division multiple access,'' \emph{Tech. Report, available online:
  \url{https://arxiv.org/abs/2002.06552}}, 2020.

\bibitem{MA2}
C.~Ciochina and H.~Sari, ``A review of \textup{OFDMA} and single-carrier
  \textup{FDMA},'' in \emph{European Wireless Conf.}\hskip 1em plus 0.5em minus
  0.4em\relax IEEE, 2010, pp. 706--710.

\bibitem{DSCDMA}
K.-C. Chen \emph{et~al.}, ``Frequency-domain approach to multiuser detection in
  \textup{DS-CDMA} communications,'' \emph{IEEE Commun., Lett.}, vol.~4,
  no.~11, pp. 331--333, 2000.

\bibitem{Allocation1}
T.~Frank, A.~Klein, E.~Costa, and E.~Schulz, ``Interleaved orthogonal frequency
  division multiple access with variable data rates,'' in \emph{Proc. Int. OFDM
  Workshop}.\hskip 1em plus 0.5em minus 0.4em\relax Citeseer, 2005, pp.
  179--183.

\bibitem{Allocation2}
R.~Dinis, D.~Falconer, C.~T. Lam, and M.~Sabbaghian, ``A multiple access scheme
  for the uplink of broadband wireless systems,'' in \emph{GLOBECOM},
  vol.~6.\hskip 1em plus 0.5em minus 0.4em\relax IEEE, 2004, pp. 3808--3812.

\bibitem{FFT2}
S.~G. Johnson and M.~Frigo, ``Implementing \textup{FFT}s in practice,''
  \emph{Fast Fourier Transforms}, 2008.

\bibitem{SwitchingBook}
S.~C. Liew and T.~T. Lee, \emph{Principles of Broadband Switching and
  Networking}.\hskip 1em plus 0.5em minus 0.4em\relax John Wiley \& Sons, 2010,
  vol.~32.

\bibitem{BP1}
P.~Schniter, ``A message-passing receiver for \textup{BICM-OFDM} over unknown
  clustered-sparse channels,'' \emph{IEEE J. Sel. Topics Signal Process.},
  vol.~5, no.~8, pp. 1462--1474, 2011.

\bibitem{BP2}
J.~S. Yedidia, W.~T. Freeman, and Y.~Weiss, ``Generalized belief propagation,''
  in \emph{Adv. Neural Inf. Process. Syst.}, 2001, pp. 689--695.

\bibitem{PAPR1}
D.~Wulich and L.~Goldfeld, ``Bound of the distribution of instantaneous power
  in single carrier modulation,'' \emph{IEEE Trans. Wireless Commun.,}, vol.~4,
  no.~4, pp. 1773--1778, 2005.

\bibitem{PAPR2}
H.~Ochiai, ``On instantaneous power distributions of single-carrier
  \textup{FDMA} signals,'' \emph{IEEE Wireless Commun. Lett.}, vol.~1, no.~2,
  pp. 73--76, 2012.

\end{thebibliography}
